  \documentclass[final,authoryear,3p,times,twocolumn]{article}
\pdfoutput=1
\usepackage{amssymb}
\usepackage{multirow}
\usepackage{rotating}
\usepackage{graphicx}
\usepackage{amsfonts}
\usepackage{amssymb}
\usepackage{natbib}
\usepackage{setspace}
\usepackage[left=1.2cm, right=1.20cm, top=1.50cm, bottom=1.50cm]{geometry}
\usepackage{hyperref}
\hypersetup{
bookmarks=false,        % show bookmarks bar?
unicode=false,          % non-Latin characters in Acrobat’s bookmarks
pdftoolbar=true,        % show Acrobat’s toolbar?
pdfmenubar=true,        % show Acrobat’s menu?
pdffitwindow=false,     % window fit to page when opened
pdfstartview={FitH},    % fits the width of the page to the window
pdftitle={ },    % title
pdfauthor={Susarla Raghuram},   % author
pdfsubject={ },   % subject of the document
pdfkeywords={ },     % list of keywords
pdfnewwindow=true,      % links in new window
colorlinks=true,        % false: boxed links; true: colored links
linkcolor=blue,         % color of internal links
citecolor=blue,        % color of links to bibliography
filecolor=magenta,      % color of file links
urlcolor=red            % color of external links
}
\newcommand{\hale}{C/1995 O1 Hale-Bopp}
\newcommand{\hbop}{Hale-Bopp}

\newcommand{\hyak}{C/1996 B2 Hyakutake}
\newcommand{\hya}{Hyakutake}

\title{Model for  Atomic Oxygen Visible Line Emissions in Comet C/1995~O1~Hale-Bopp}
\author{Susarla Raghuram\thanks{raghuramsusarla@gmail.com}
\ and Anil Bhardwaj\thanks{anil\_bhardwaj@vssc.gov.in;
bhardwaj\_spl@yahoo.com} \\ Space Physics Laboratory, 
Vikram Sarabhai Space Center, Trivandrum, 695022, India.}

\begin{document}
\maketitle
% \begin{frontmatter}



% \title{Model for  Atomic Oxygen Visible Line Emissions in Comet C/1995~O1~Hale-Bopp}
% 
% %% use optional labels to link authors explicitly to addresses:
% \author{Susarla Raghuram}
% \ead{raghuramsusarla@gmail.com}
% \author{Anil Bhardwaj\corref{cor1}} %\fnref{lable2}}
% \cortext[cor1]{Corresponding author. Fax: +91 471 2706535 } %: anil\_bhardwaj@vssc.gov.in; bhardwaj\_spl@yahoo.com}
% \ead{Anil\_Bhardwaj@vssc.gov.in; bhardwaj\_spl@yahoo.com}
% \address{Space Physics Laboratory,
% Vikram Sarabhai Space Centre,
% Trivandrum~695022, India}

% \address{Space Physics Laboratory, 
% Vikram Sarabhai Space Center, Trivandrum 695022, India.}

\begin{abstract}
 We have recently developed a coupled chemistry-emission model for the green
 (5577 \AA) and red-doublet 
(6300, 6364 \AA) emissions of atomic oxygen on comet \hyak. In the present work
we applied our model  to comet \hale,  which had  an order of magnitude
 higher H$_2$O production rate  than comet Hyakutake,
to evaluate the photochemistry  associated with the production and loss of 
O($^1$S) and O($^1$D) atoms and emission processes of green and red-doublet lines.  
We present the wavelength-dependent photo-attenuation rates for different 
photodissociation processes forming O($^1$S) and O($^1$D).
 The calculated radiative efficiency profiles of O($^1$S) and O($^1$D) 
atoms  show that  in comet \hbop\ the green and red-doublet emissions are emitted  
mostly  above radial distances of 10$^3$ and  10$^4$ km, respectively. 
The model calculated  [OI] 6300 \AA\ emission surface brightness and  average intensity over 
the Fabry-P{\'e}rot spectrometer field of view  are consistent with the observation of
\cite{Morgenthaler01}, while the intensity ratio of green to red-doublet emission 
 is in agreement with the  observation of \cite{Zhang01}. 
In comet \hbop, for cometocentric distances less than 10$^{5}$ km, the intensity 
of [OI]  6300 \AA\ line is  mainly governed by photodissociation of H$_2$O. Beyond 10$^{5}$ km,
O($^1$D) production is dominated by photodissociation of the water photochemical daughter 
product OH. Whereas the [OI] 5577 \AA\ emission line is controlled by photodissociation 
of both H$_2$O and CO$_2$.
 The calculated mean excess energy in various photodissociation 
processes show that the photodissociation of CO$_2$ can produce O($^1$S) atoms with 
higher excess velocity compared to the photodissociation of H$_2$O.
 Thus,
our model calculations suggest that involvement of multiple sources in the formation of
O($^1$S) could be a reason for the larger width of green line than that of
red-doublet emission lines observed in several comets.
\end{abstract}
% 
% \begin{keyword}
% Comets \sep Comets coma \sep Comets composition \sep Comet Hale-Bopp \sep Comet Hyakutake \sep
% Photochemistry \sep Spectroscopy
% %% keywords here, in the form: keyword \sep keyword
% 
% %% MSC codes here, in the form: \MSC code \sep code
% %% or \MSC[2008] code \sep code (2000 is the default)
% 
% \end{keyword}

% \end{frontmatter}

%  \linenumbers

\section{Introduction}
Owing to its very high H$_2$O production rate, \hale\ was a
great comet in the night sky of the year 1997. The visible 
emissions of atomic oxygen ([OI] 6300, 6364,  and 5577 \AA), which are accessible 
to the ground-based optical instruments, have been observed by \cite{Morgenthaler01}
 and \cite{Zhang01} in the coma of \hbop. Since the lifetime of oxygen atom
 in the $^1$D  metastable state is relatively small ($\sim$110 s) compared to the 
photochemical lifetime of H$_2$O ($\sim$8 $\times$ 10$^4$ s), it cannot 
travel larger distances in the coma without decaying to the ground $^3$P state. Moreover, 
most of the production of oxygen in the $^1$D state is through photodissociative
 excitation of H$_2$O \citep{Bhardwaj02}. Thus, [OI] 6300 \AA\ 
emission has been used to 
trace the spatial distribution as well as to quantify the production rate of H$_2$O in 
 several comets \citep{Delsemme76,Delsemme79,Fink84,
Schultz92,Morgenthaler01,Furusho06,Fink09}.

Based on the  study of \cite{Festou81}
the intensity ratio of green (5577 \AA) to red-doublet (6300 \AA\ and 6364 \AA) emissions 
(here after G/R ratio) of atomic oxygen has been used to determine whether the parent source of 
these lines is H$_2$O  or CO$_2$/CO in the  coma of comets \citep{Cochran84,Cochran08,
Morrison97,Zhang01,Cochran01,Furusho06,Capria05,Capria08,Capria10,McKay12,McKay12b}. The modelling 
studies of these emissions in comets showed that photodissociative excitation of H$_2$O is the major
 production process of the [OI] 6300 \AA\ emission \citep{Festou81,Bhardwaj02,Capria05,Capria08,Bhardwaj12}. 
Our recent theoretical study  \citep{Bhardwaj12} for these prompt emissions of atomic oxygen 
in comet \hyak\ showed that 
 more than 90\% of the O($^1$D) is populated via photodissociative excitation of H$_2$O and the
rest through photodissociation of  other oxygen bearing species, like CO$_2$ and CO. It also 
showed that quenching  by H$_2$O is the major loss mechanism of O($^1$D) up to  
cometocentric distances of 1000 km, and above that distance radiative decay takes over. 
The study of \cite{Bhardwaj12}
demonstrated that the G/R ratio depends  not only on the photochemistry involved in populating 
O($^1$S) and O($^1$D) atoms in the 
cometary coma,  but also on the projected area observed for the comet, which is a 
function of slit 
dimension used for observation and geocentric distance of the comet.

In the present study we applied our coupled chemistry-emission model \citep{Bhardwaj12}  to  
comet \hale, which had an 
 order of magnitude higher H$_2$O production rate 
 compared to that of comet Hyakutake, to evaluate the production and 
loss mechanisms of O($^1$D) and O($^1$S) 
and generation of green and red-doublet emissions. Our aim in this paper is to study the photo-attenuation
in comets having high H$_2$O production rates and its implications on the photochemistry of 
metastable O($^1$S) and O($^1$D) atoms.
 We compared our model calculated [OI] 6300 \AA\ emission surface brightness profile
with the observation of \cite{Morgenthaler01}. 
We have shown that the photodissociation of H$_2$O  mainly controls 
 the formation of O($^1$D) and subsequently determines the intensity of  [OI] 6300 \AA\ emission.
 However, in the case of [OI]  5577 \AA\ emission, the photodissociation of both H$_2$O and CO$_2$ 
plays an important role in the formation of metastable O($^1$S), with  photodissociation of CO$_2$ 
being the major production source of O($^1$S) in the inner cometary coma.  We suggest that 
in comets with significant ($\ge$5\%)  CO$_2$ 
relative abundance, the photodissociation of CO$_2$  is more effective in populating 
O($^1$S) than the photodissociation of H$_2$O. The model calculated G/R ratio is consistent with the 
observed value of \cite{Zhang01}. Based on the 
model results, we suggest that the involvement of  multiple species in the formation O($^1$S) 
could be  a reason for the width of the green line emission to be larger than the 
red-doublet emission lines observed in several comets by \cite{Cochran08}.

\section{Model}
\label{sec:model}
The details of the  model and the chemistry are presented in our 
previous work \citep{Bhardwaj12}. Here we present the input parameters that 
have been used in the model for the observed condition of comet Hale-Bopp on 
10 March 1997 (solar radio flux 
F10.7 = 74.7 $\times$ 10$^{-22}$ J s$^{-1}$ m$^{-2}$ Hz$^{-1}$)
when it was at a geocentric distance ($\Delta$) of 1.32 AU and a heliocentric distance 
($r_h$) of 0.93 AU. The photochemical reaction network and cross sections for photon and 
electron impact processes are same as used in the previous work and any changes made are 
mentioned. The degradation of solar EUV-generated photoelectrons is 
accounted by using Analytical Yield Spectrum (AYS) technique which is based on the Monte-Carlo 
method \citep{Singhal91,Bhardwaj93,Bhardwaj99d,Bhardwaj99b,Bhardwaj09}. Details of the  AYS approach
and the method to calculate photoelectron flux and excitation rates are given in 
our earlier papers \citep{Bhardwaj90,Bhardwaj96,Bhardwaj99a,Bhardwaj03,Haider05,Bhardwaj11,
Raghuram12,Bhardwaj12c}.

 \begin{center} 
\begin{table*}[tbh]  % Table 1
\small
  \caption{Major production and destruction processes of the  O($^1$S) and  O($^1$D). Photorates 
 are calculated using solar flux on 10 Apr 1997 (solar minimum period : solar radio flux 
F10.7 = 74.7 $\times$ 10$^{-22}$ J s$^{-1}$ m$^{-2}$ Hz$^{-1}$.) and scaled to 0.92 
AU heliocentric distance.}
\renewcommand{\thefootnote}{\fnsymbol{footnote}}
 \begin{center} 
\scalebox{0.8}[0.8]{
  \begin{tabular}{llll|l|l|l|l|l|l|l|l|l|l|l|}
 \hline
  Reaction & Rate (cm$^{-3}$ s$^{-1}$ or s$^{-1}$)& Reference\\
\hline  
H$_2$O + h$\nu$ $\rightarrow$ O($^1$S) + H$_2$ 
& 3.78 $\times$ 10$^{-8}$  & This work \\  
OH + h$\nu$  $\rightarrow$ O($^1$S) + H
& 6.71 $\times$ 10$^{-8}$  &\cite{Huebner92}\footnotemark[3]\\  
CO$_2$  + h$\nu$ $\rightarrow$ O($^1$S) + CO
& 8.5 $\times$ 10$^{-7}$ & This work\\ 
CO  + h$\nu$ $\rightarrow$  O($^1$S) + C
& 4.0 $\times$ 10$^{-8}$ & \cite{Huebner79}  \\
H$_2$O$^+$ + e$_{th}$ $\rightarrow$ O($^1$S) + others
& 4.3 $\times$ 10$^{-7}$ (300/T$_e$)$^{0.5}$ $\times$ 0.045\footnotemark[2] 
& \cite{Rosen00}\\
O($^1$S) + H$_2$O $\rightarrow$ 2 OH
& 3 $\times$ 10$^{-10}$ & \cite{Zipf69}\\  
O($^1$S) \hspace{0cm} $\longrightarrow$ O($^3$P) + h$\nu_{2972\AA}$
& 0.134  & \cite{Slanger06} \\  
O($^1$S) \hspace{0cm} $\longrightarrow$ O($^1$D) + h$\nu_{5577\AA}$
& 1.26  & \cite{Wiese96}\\   
 H$_2$O + h$\nu$ $\rightarrow$ O($^1$D) + H$_2$ & 9.5  $\times$ 10$^{-7}$& This work \\ 
OH + h$\nu$ $\rightarrow$ O($^1$D) + H 
& 7.01 $\times$ 10$^{-6}$ & \cite{Huebner92}\footnotemark[4]\\  %
CO$_2$ + h$\nu$ $\rightarrow$ O($^1$D) + CO    &6.2 $\times$ 10$^{-7}$ & This work\\   
CO + h$\nu$ $\rightarrow$ O($^1$D) + C         &6.0  $\times$ 10$^{-8}$ & This work\\	
H$_2$O$^+$ + e$_{th}$ $\rightarrow$ O($^1$D) + others
& 4.3 $\times$ 10$^{-7}$ (300/T$_e$)$^{0.5}$ $\times$ 0.045\footnotemark[2] 
& \cite{Rosen00}\\
CO$^+$ + e$_{th}$ $\rightarrow$ O($^1$D) + others
&5.0 $\times$ 10$^{-8}$ $\times$ (300/T$_e$)$^{0.46}$ & \cite{Mitchell90} \\  
O($^1$D) + H$_2$O  $\rightarrow$ 2 OH
& 2.1 $\times$ 10$^{-10}$ & \cite{Atkinson97}\\  
O($^1$D) \hspace{0cm} $\longrightarrow$ O($^3$P)+ h$\nu_{6300\AA}$ 
& 6.44 $\times$ 10$^{-3}$ & \cite{Storey00}\\   
O($^1$D) \hspace{0cm} $\longrightarrow$ O($^3$P)+ h$\nu_{6364\AA}$ 
&2.15 $\times$ 10$^{-3}$ & \cite{Storey00}\\  
 \hline 
  \end{tabular}}
 \end{center}
\footnotemark[2]{0.045 is the assumed branching ratio for the formation of 
O($^1$S) and O($^1$D) via dissociative recombination of H$_2$O$^+$ ion \citep[see][]{Bhardwaj12}.}
\footnotemark[3]{\cite{Huebner92} calculated this rate using theoretical
 OH absorption cross section of \cite{Dishoeck84}.}
\footnotemark[4]{\cite{Huebner92} calculated this rate based on 
experimentally determined OH absorption cross section of \cite{Nee84}.}
 {h$\nu$ : solar photon;  e$_{th}$ : thermal
electron;  T$_e$ : electron temperature.} 
\label{tabo1sd}
\end{table*} 
\end{center}

In our previous work \citep{Bhardwaj12} it has been shown that the contribution of several 
processes to the production of O($^1$S) and O($^1$D)  is
  small. Thus, only important production and destruction mechanisms 
 of metastable O($^1$S) and O($^1$D) are presented in Table~\ref{tabo1sd}.
 The branching ratio of 0.81 is used for the production of OH
in the photodissociation of H$_2$O \citep[cf.][]{Huebner92,Nee84}. The radius of the cometary
nucleus is taken as 25 km \citep{Weaver97,Combi99}. 
Though a variety of O-bearing minor species \citep[with relative abundances $\le$1\%,][]{Bockelee00} 
have been found in comet \hbop, the dominant species  
H$_2$O, CO$_2$, and CO are only considered in our model calculations. The neutral density profiles 
of these parent species are  calculated using Haser's formula. 

Using ground-based  
observations of direct H$_2$O infrared emissions during January to May 1997, \cite{Russo00} 
derived water production rates 
at different heliocentric distances and fitted  
as a function of $r_h$ as Q$_{H_2O}$ = 8.35 $\pm$  0.13 $\times$ 10$^{30}$ [$r_h^{(-1.88 \pm  0.13)}$] 
molecules s$^{-1}$. We used this expression as a standard input in calculating  H$_2$O density profiles
on different days.
The H$_2$O production rates in \hbop\ are also derived by observing the emissions from its dissociative 
products, like OH 18-cm emission, OH (0-0) 3080 \AA\ emission, [OI] 6300 \AA\ emission, and
  H Lyman-$\alpha$, over large projected distances
 \citep{Weaver97,Colom99,Schleicher97,Combi00,Woods00,Morgenthaler01,Harris02,Fink09}. These derived 
H$_2$O production rates depend on the observational condition and also on the assumed model parameters.
 We have discussed the effect of H$_2$O production rate on the 
calculated green and red-doublet emission intensities of atomic oxygen in the Section \ref{relabn}.

High resolution ground-based infrared spectroscopic observations 
are used to study the CO production 
rate in comet \hbop\ from June 1996 to September 1997  \citep{Disanti01}.
 The spatial distribution of CO
in the coma of \hbop\ is assumed to have two distinct sources: 
 nucleus-originated, 
and extensively distributed in the cometary coma. 
During this  observation period, 
 the relative abundance of CO 
 is found to be $\sim$25\% of  H$_2$O. \cite{Disanti01} fitted  the observed   CO production 
rate as a function of heliocentric distance near the perihelion as 
Q$_{CO}$~= 2.07 $\times$ 10$^{30}$ r$_{h}^{-1.66\pm0.22}$ molecules s$^{-1}$, and above 1.5 AU as 
Q$_{CO}$~= 1.06 $\times$ 10$^{30}$ r$_{h}^{-1.76\pm0.26}$ molecules s$^{-1}$. 
Since observations of [OI] 6300 \AA\ emission are done when comet was at around 1 AU,
we have used the former formulation to calculate 
the neutral CO density in our model. \cite{Disanti01} suggested that the increase in CO production 
rate below 1.5 AU is due to distributed sources prevailing in the cometary coma.
Recent study of \cite{Bockelee10} showed that 
the infrared CO (1-0) rotational vibrational emission lines are optical 
thick in the cometary coma of \hbop. Based on the modelling studies of these emission lines 
they rejected the idea of
 extended source distribution of CO in comet \hbop. However, our model calculations show that
 the role of CO 
in determining green and red-doublet emission intensities is very small compared to other species,
and hence the impact of distributed CO source is insignificant on these forbidden emission
lines.

The CO$_2$  has been detected in \hbop\ by \cite{Crovisier97}  in April 1996, when  
 the comet was at heliocentric distance of 2.9 AU.  
Based on the infrared emissions between 2.5 to 5 $\mu$m, the derived  CO$_2$ production rate 
at 2.9 AU was 1.3 $\times$ 10$^{28}$ 
molecules s$^{-1}$, which corresponds to  a relative abundance of $\sim$20\% of H$_2$O.
Assuming that the photodissociative excitation is the
main production mechanism in populating the CO(a$^3\Pi$) metastable state,
the observed CO Cameron band (a$^3\Pi$ $\rightarrow$ X$^1\Sigma^+$) emission  
intensity has been 
used to estimate the abundance of CO$_2$ in this comet by \cite{Weaver97}. 
The estimated CO$_2$ abundance is more than 10\% when the comet was beyond 2.7 AU. However, 
our model calculations  on comets 103P/Hartley 2 \citep{Bhardwaj11} and 1P/Halley \citep{Raghuram12} 
have shown that photoelectron impact 
excitation is the main production mechanism of CO Cameron band emission and not the 
photodissociation of CO$_2$.
Assuming that the CO$_2$/CO abundance ratio did not vary with heliocentric distance in this comet, 
\cite{Bockelee04} suggested 6\% relative abundance of CO$_2$ when the comet was 
at 1 AU. We have taken 6\% CO$_2$ relative abundance with respect 
to H$_2$O in the model.
However, we discuss  the 
impact of CO$_2$ abundance by varying its relative abundance on the calculated intensities of green and 
red-doublet emissions.
The OH neutral density profile in comet Hale-Bopp is calculated by 
 fitting \cite{Harris02} observed  OH (0-0) 3080 \AA\ resonant scattering emission 
along the projected distance with the Haser's two step formulation. The photodissociative
excitation rates of OH producing O($^1$S) and O($^1$D) are taken from \cite{Huebner92} which 
were determined using theoretical \citep{Dishoeck84} and experimental \citep{Nee84}
photoabsorption cross sections, respectively.

There is a clear evidence that in comet \hbop\ the expansion velocity of neutrals 
 increases with increasing cometocentric
distance \citep{Colom99,Biver97,Harris02}. The sources involved in accelerating 
the neutral species across the cometary coma is discussed in several works 
\citep{Colom99,Combi99,Harris02,Combi02}. 
To incorporate the acceleration of these neutrals in our model we have taken the velocity profile 
calculated by \cite{Combi99} at 1 AU and used as a input in the Haser's density distribution. 
We also verified the effect of expansion velocity on
the calculated intensity of green and red-doublet emissions by varying its static value between
0.7 to 2.2 km s$^{-1}$, which is discussed in Section~\ref{eff-vel}.
\begin{center}
\begin{figure}
\noindent\includegraphics[width=22pc,angle=0]{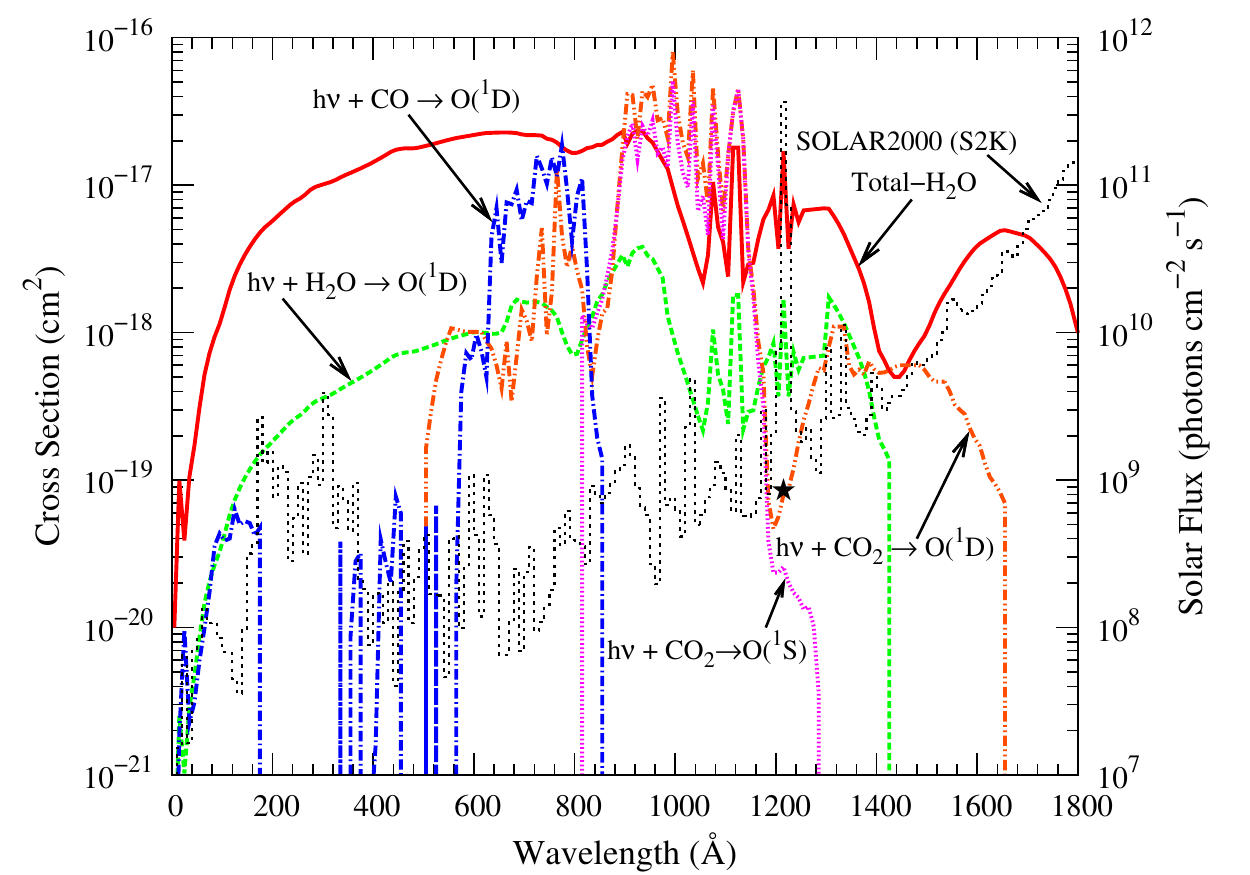}
\caption{Photodissociative excitation cross sections for the production of O($^1$D) from H$_2$O 
and CO  are taken from \cite{Huebner92}. The photodissociation of CO$_2$ for the 
production of O($^1$D) are taken from  \cite{Jain12}. 
The photodissociation cross section 
of CO$_2$ producing O($^1$S) is calculated using the yield suggested by \cite{Huestis10} and 
total absorption cross section.  $\bigstar$
represents the cross section value for the production of O($^1$S) from H$_2$O at 1216 \AA\ 
assuming 0.5\% yield.
For comparison the total photoabsorption cross section of H$_2$O taken from \cite{Huebner92} 
is also shown. 
The solar flux taken from SOLAR2000 (S2K) model on 9 March 1997 (solar minimum condition; 
solar radio flux 
F10.7 = 74.7 $\times$ 10$^{-22}$ J s$^{-1}$ m$^{-2}$ Hz$^{-1}$)
 is shown with scale on the right side y-axis.}
\label{o-csc}
\end{figure}
\end{center}

The input solar flux is taken from SOLAR2000 (S2K) v.2.36 model of \cite{Tobiska04} 
 and scaled accordingly to the heliocentric distance of  the comet 
at the time of observation. 
The electron temperature profile required to calculate dissociative 
recombination rates is taken from  \cite{Lovell04}. 
\cite{Bhardwaj12} have found that the  yield of O($^1$S) in the photodissociation of H$_2$O at solar
H Ly-$\alpha$ can not be more than 1\%. In the present study we have taken this yield value as 0.5\%.
The impact of this assumption was discussed in our previous work \citep{Bhardwaj12}.
The photodissociative excitation cross section for CO$_2$ producing  O($^1$D) is taken
from \cite{Jain12}. The photodissociative excitation cross sections 
for the production of O($^1$D) 
and O($^1$S) from H$_2$O, CO$_2$, and CO used in the model are presented in Figure~\ref{o-csc}.
The attenuation of solar radiation and solar UV-EUV generated photoelectrons in the cometary 
coma are described in our previous works \citep{Bhardwaj90,Bhardwaj99a,Bhardwaj03, 
Bhardwaj99c,Raghuram12}. 

\begin{figure}
\begin{center}
 \noindent\includegraphics[width=22pc,angle=0]{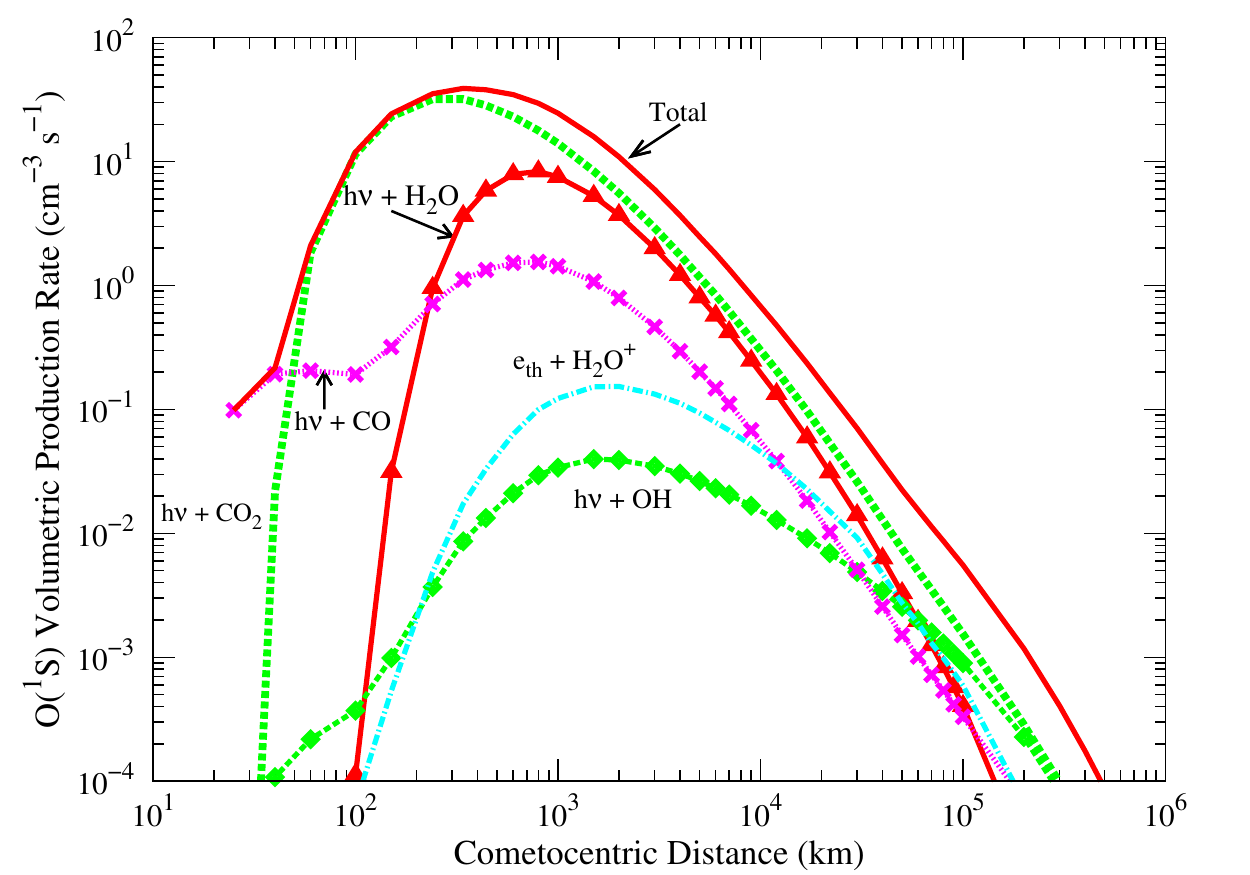}
\caption{Calculated radial profiles for major production mechanisms of O($^1$S) along with the 
total production profile for the abundances of 6\%  CO$_2$ and 24\%  CO relative to H$_2$O production
rate of 8.3 $\times$ 10$^{30}$ s$^{-1}$.
h$\nu$ : solar photon and e$_{th}$ : thermal electron}
\label{prat-o1s}
\end{center}
\end{figure}

\section{Results} 
\subsection{Production and loss of O($^1$S)}
The calculated  O($^1$S) volumetric production rate profiles for major production processes are presented in 
Figure~\ref{prat-o1s}. The photodissociation of CO$_2$ is the major production process  of O($^1$S). 
Above  cometocentric distance  of 1000 km, the photodissociative excitation of H$_2$O is also an equally 
important production source of O($^1$S). Photodissociative excitation of CO is the next 
significant production 
mechanism in producing O($^1$S). Since no cross section is reported in the literature for photodissociation 
of CO producing O($^1$S), we have taken the photo-rate for this process from \cite{Huebner79} 
and assumed that the formation of O($^1$S) is similar to O($^1$D).
This assumption results in the calculated O($^1$S) profile  
below 100 km  similar to that of O($^1$D). 
However, this assumption does not make any significant impact on the calculated green line intensity, 
 since photodissociation of CO$_2$ and H$_2$O can produce O($^1$S) an order  
of magnitude higher than that of CO in the inner coma.
Above 10$^4$ km,  the contribution from dissociative recombination reactions of H$_2$O$^+$ and CO$^+$
  to the total O($^1$S) production is significant. The photodissociation of OH is a
minor source of O($^1$S) below 10$^5$ km radial distance.

\begin{figure}
\begin{center}
\noindent\includegraphics[width=22pc,angle=0]{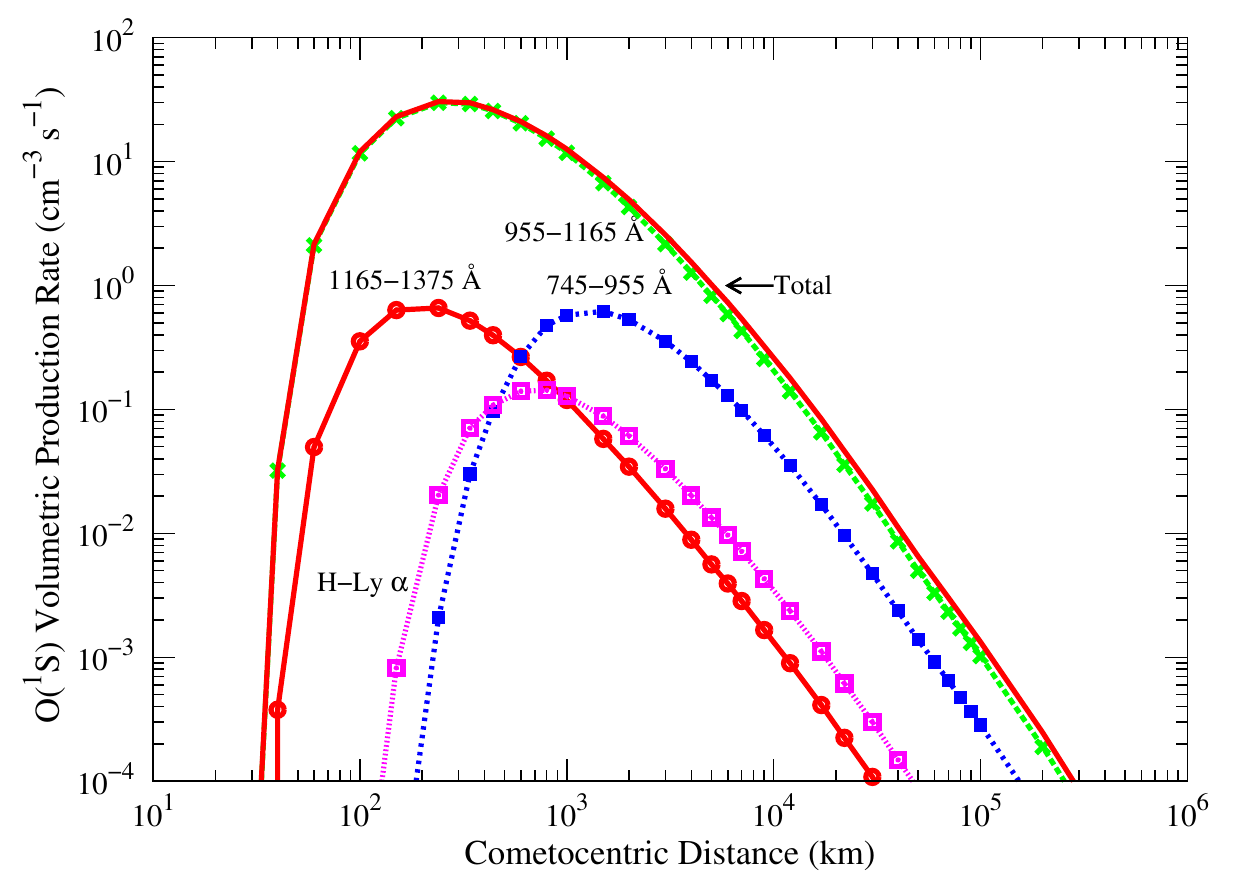}
\caption{Calculated radial profiles for the photodissociation of CO$_2$ producing O($^1$S) 
at different wavelength bands for the abundances of 6\% CO$_2$ and 24\%  CO relative to H$_2$O
 production
rate of 8.3 $\times$ 10$^{30}$ s$^{-1}$.}
\label{wav-o1s-co2}
\end{center}
\end{figure}

The calculated O($^1$S) volumetric production rate profiles for photodissociation of CO$_2$ in the 
different wavelength bands are shown in Figure~\ref{wav-o1s-co2}.  The cross section for 
photodissociation of CO$_2$ in the wavelength band 955--1165 \AA\ is higher by a 
few orders of magnitude compared to that at other wavelength regions  (cf.~Fig.~\ref{o-csc}). 
Moreover, in this wavelength band, the
 yield of O($^1$S) in photodissociation of CO$_2$ tends to unity \citep{Slanger77,Lawrence72}, while 
the total absorption cross section of H$_2$O has a strong dip  (cf.~Fig.~\ref{o-csc}).
Thus, solar photons in this wavelength band can dissociate CO$_2$ and produce O($^1$S) very 
efficiently. The photons in the wavelength bands 1165-1375 and 745-955 \AA\ make a smaller ($<$10\%)
contribution to the total O($^1$S) production. The contribution of 1216  \AA\ solar photons to 
the O($^1$S) formation is two orders of magnitude low because of the small absorption cross section of CO$_2$ 
($\sim$8 $\times$ 10$^{-20}$ cm$^{2}$).

\begin{figure}[h]
\begin{center}
\noindent\includegraphics[width=22pc,angle=0]{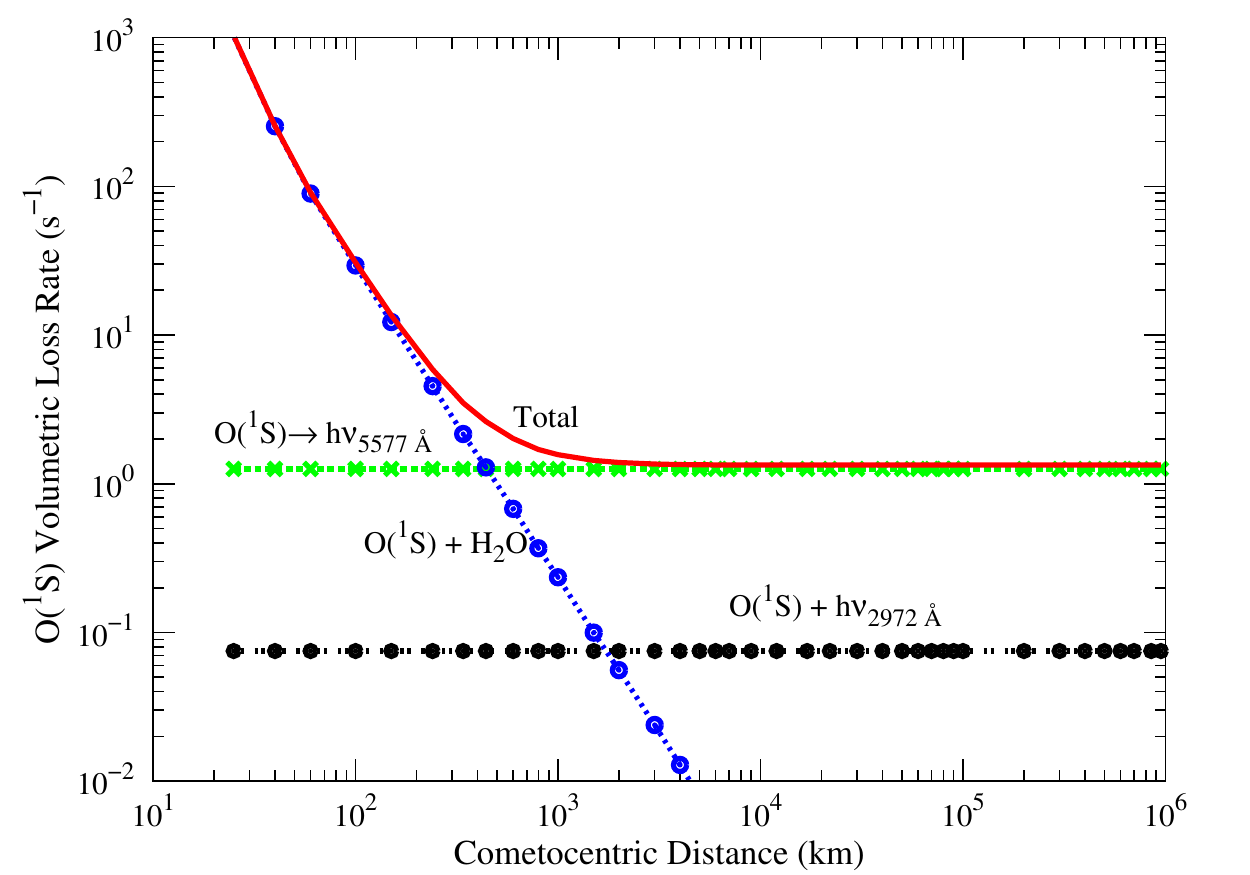}
\caption{Calculated radial profiles for the major loss mechanisms of the O($^1$S) atom
 for the abundances of 6\%  CO$_2$ and 24\%   CO relative to H$_2$O
 production
rate of 8.3 $\times$ 10$^{30}$ s$^{-1}$.}
\label{lrate-o1s}
\end{center}
\end{figure}

The calculated volumetric destruction rate profiles of O($^1$S) are presented in 
Figure~\ref{lrate-o1s}. The collisional quenching of O($^1$S) by H$_2$O is the dominant loss process 
at cometocentric distances shorter than 300 km.  Above 1000 km the radiative decay  via [OI] 5577 \AA\ 
line emission  is the main loss process for the O($^1$S) atom. The radiative decay
 via [OI] 2972 \AA\ emission is a minor loss process of O($^1$S).

\begin{figure}[h]
\begin{center}
\noindent\includegraphics[width=22pc,angle=0]{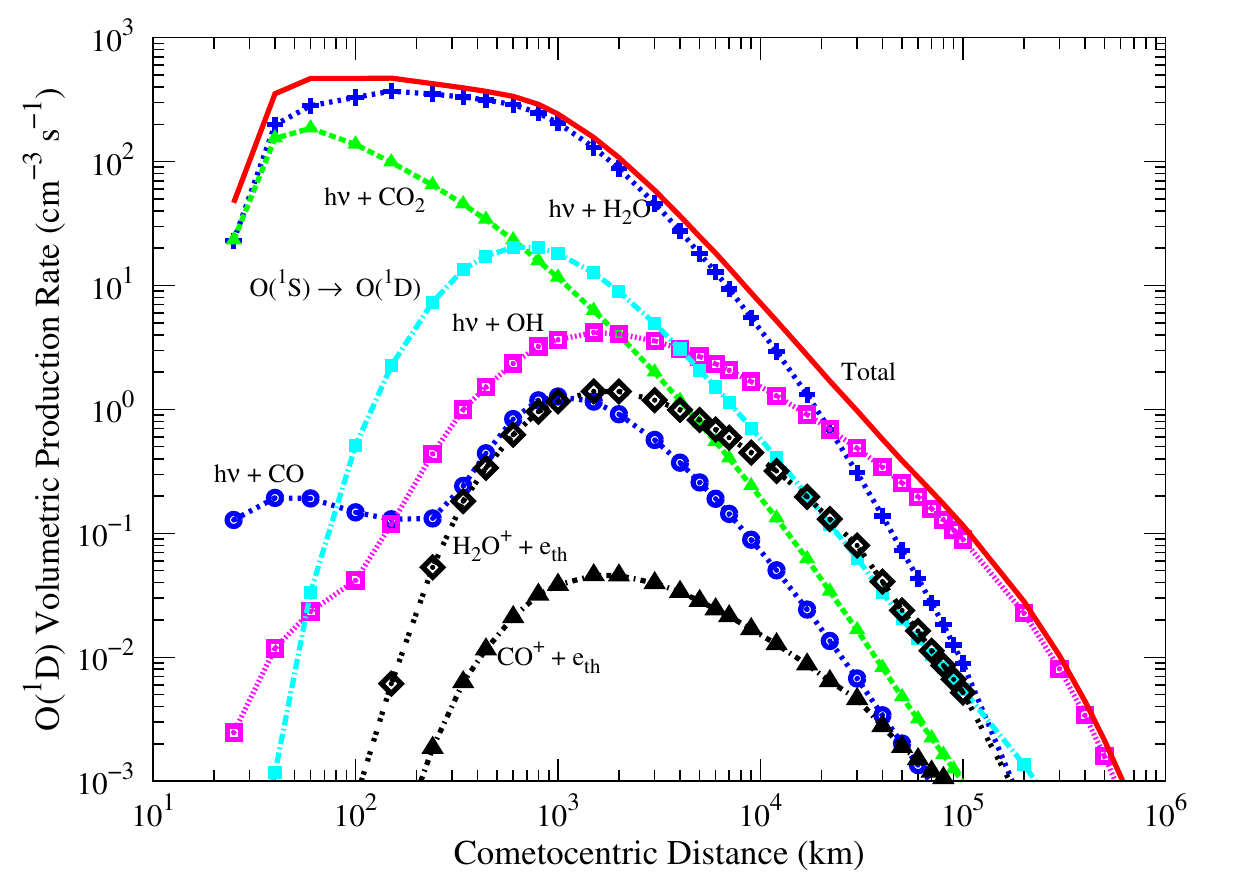}
\caption{Calculated radial profiles for the major production mechanisms of O($^1$D) 
along with the total O($^1$D) production rate profile
 for the abundances of 6\%  CO$_2$ and 24\% CO relative to H$_2$O  production
rate of 8.3 $\times$ 10$^{30}$ s$^{-1}$. h$\nu$ : solar photon.}
\label{prat-o1d}
\end{center}
\end{figure}

\subsection{Production and loss of O($^1$D)}
The calculated volumetric production rate profiles of metastable O($^1$D) 
for different formation mechanisms are shown in Figure~\ref{prat-o1d}. 
Between 100 and $\sim$2 $\times$ 10$^4$ km, most of the O($^1$D) ($>$90\%)
 is produced via photodissociation of H$_2$O. However, below 100 km, 
 the photodissociation of CO$_2$ is also an important source of  O($^1$D). 
Between 200 and 2000 km, the radiative decay of 
O($^1$S) makes a minor contribution in the formation of O($^1$D). Above 10$^4$ km, 
the photodissociation of OH plays a significant role in 
the formation of O($^1$D). Even though the relative abundance of CO in  \hbop\ 
is high ($\sim$25\%), the photodissociation of CO  is not a potential
source mechanism of O($^1$D). The calculated O($^1$D) photodissociation rate profile for 
photodissociation of CO shows a double peak structure, which is explained later.

\begin{figure}[h]
\begin{center}
\noindent\includegraphics[width=22pc,angle=0]{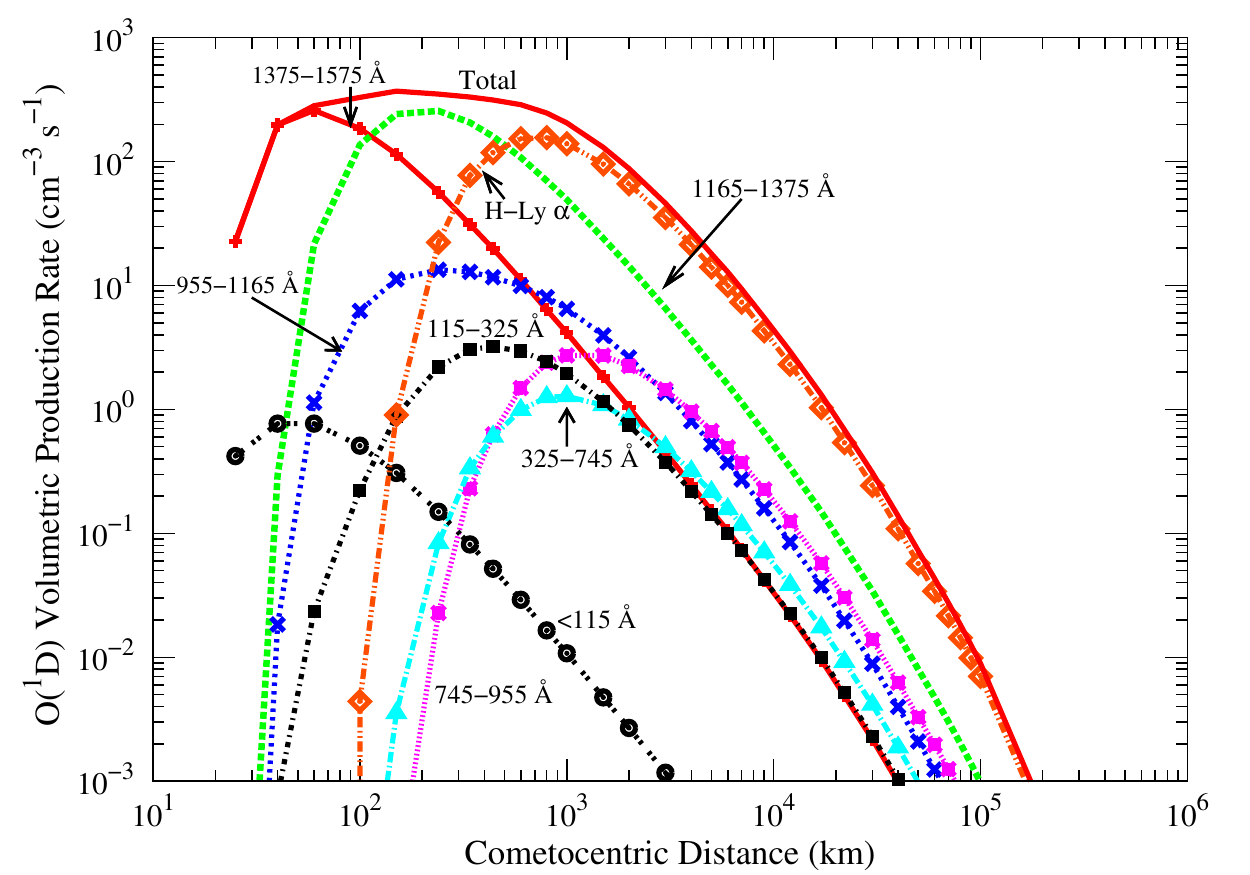}
\caption{Calculated radial profiles for the photodissociation of H$_2$O producing O($^1$D) 
at different wavelength bands  for the abundances of 6\%  CO$_2$ and 24\%  CO relative to H$_2$O
 production
rate of 8.3 $\times$ 10$^{30}$ s$^{-1}$.}
\label{wav-o1d-h2o}
\end{center}
\end{figure}

The wavelength-dependent production rates of O($^1$D) in the 
photodissociation of H$_2$O are shown in Figure~\ref{wav-o1d-h2o}.
 The most intense line of solar UV  spectrum, H Ly-$\alpha$ at 1216 \AA, produces 
maximum O($^1$D) around 1000 km, 
 while solar photons in the wavelength regions 1165--1375 
 and 1375--1575 \AA\ are responsible for producing maximum O($^1$D) at shorter radial 
distances of 200  and 50 km,  respectively.
 Since the total absorption cross section of H$_2$O in the 1165--1575 \AA\ wavelength region 
is small  (cf.~Fig.~\ref{o-csc}), these solar photons are able to penetrate deeper in the coma
 and mostly get  attenuated at shorter cometocentric distances by dissociating H$_2$O.
 The O($^1$D) formation rate
 by solar photons at other wavelengths is smaller by more than an order of magnitude.

\begin{figure}[h]
\begin{center}
\noindent\includegraphics[width=22pc,angle=0]{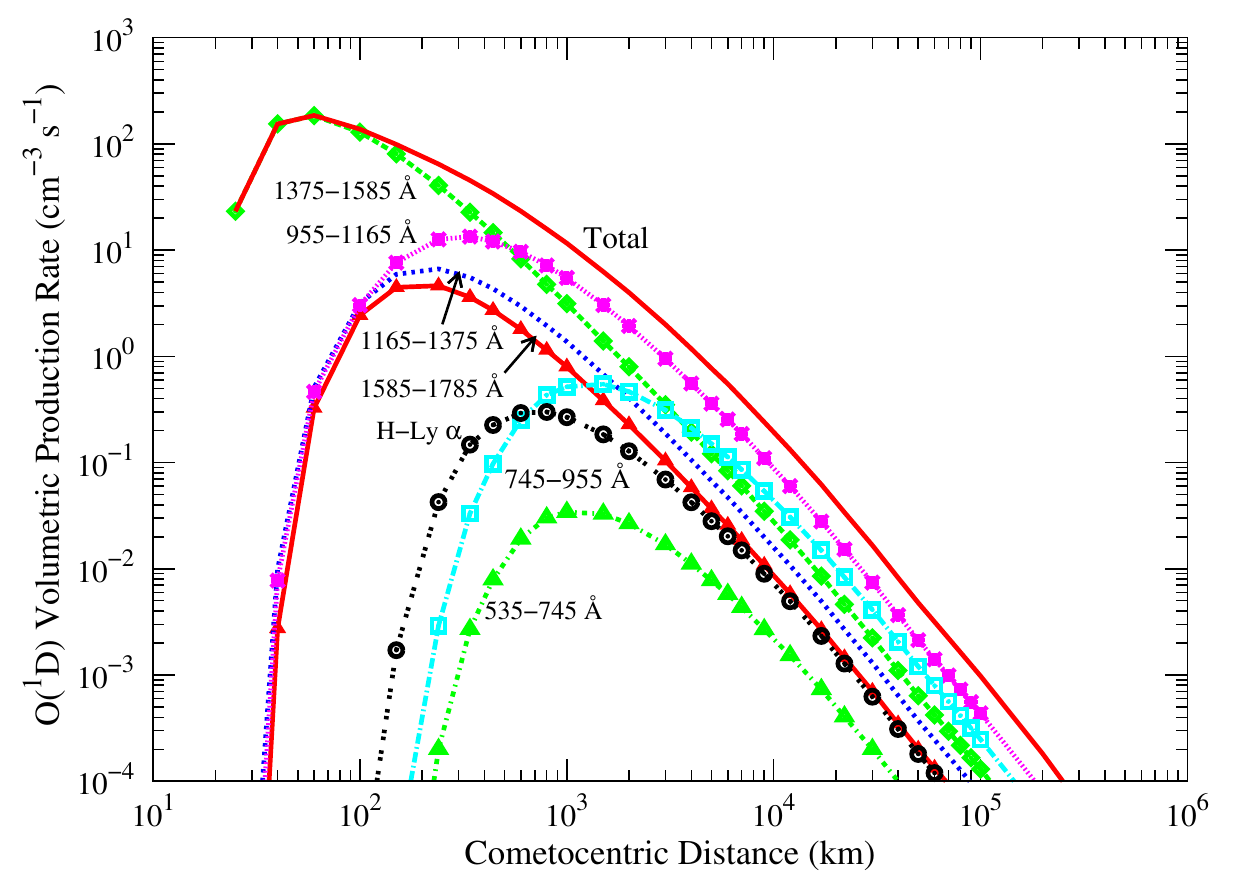}
\caption{Calculated radial profiles for the photodissociation of CO$_2$ producing O($^1$D) 
at different wavelength bands  for the abundances of 6\%  CO$_2$ and 24\% CO relative to H$_2$O
 production
rate of 8.3 $\times$ 10$^{30}$ s$^{-1}$.}
\label{wav-o1d-co2}
\end{center}
\end{figure}
Similarly, the production rate of O($^1$D) due to  
photodissociation of CO$_2$ calculated at  different wavelength bands is shown in 
Figure~\ref{wav-o1d-co2}. At radial distances $<$100 km, solar photons in 
1375--1585 \AA\  wavelength region is the main source for O($^1$D) formation. 
This is because the absorption cross section 
of H$_2$O has a strong dip around 1400 \AA\ (cf.~Fig.~\ref{o-csc}) and
the average absorption cross section values of H$_2$O and CO$_2$ are nearly same 
in this wavelength region. 
Thus, solar photons in this wavelength band are able to 
reach the innermost coma and produce O($^1$D) by dissociating CO$_2$.
Since the cross section    
for production of O($^1$D) in photodissociation of CO$_2$ peaks in the wavelength band 955--1165 \AA,
 the solar photons of this region leads the production of  O($^1$D) above 500 km.

\begin{figure}[h]
\begin{center}
\noindent\includegraphics[width=22pc,angle=0]{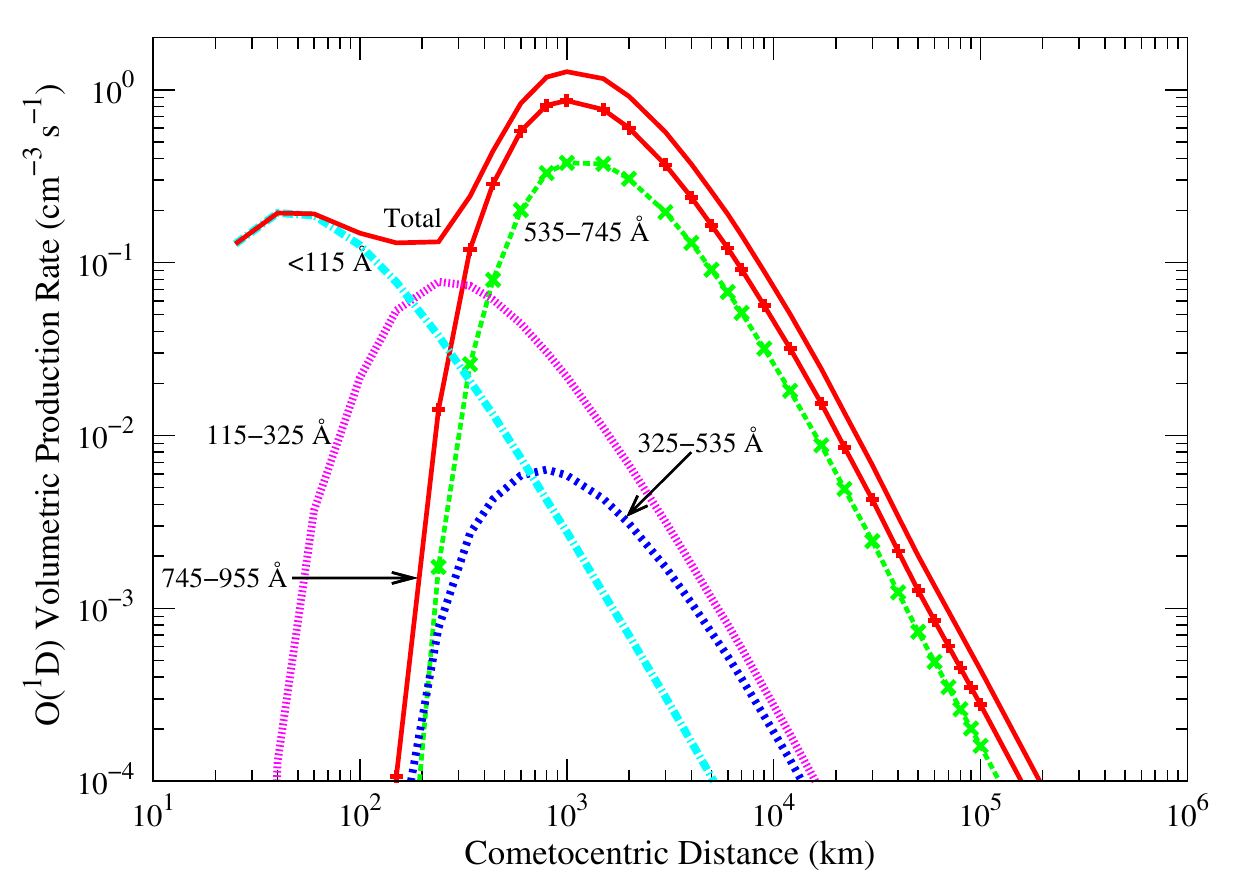}
\caption{Calculated radial profiles for the photodissociation of CO producing O($^1$D) 
at different wavelength bands  for the abundances of 6\%  CO$_2$ and 24\% CO relative to H$_2$O
 production
rate of 8.3 $\times$ 10$^{30}$ s$^{-1}$.}
\label{wav-o1d-co}
\end{center}
\end{figure}

 The production rates of O($^1$D) via photodissociation of CO in different wavelength bands 
are presented in Figure~\ref{wav-o1d-co}. 
The total absorption cross section of H$_2$O is around two orders of magnitude smaller below 
115 \AA\ than at other wavelengths, so these high energy photons can travel deeper into the 
cometary coma (even below 100 km) almost unattenuated. 
Since the CO molecule offers a  cross section (average $\sim$2 $\times$ 
10$^{-20}$ cm$^2$) to these photons it leads to the
 formation of O($^1$D) and C($^1$D) via photodissociation closer to the cometary nucleus. 
Between 100 and 500 km,
 the solar photons in the wavelength region 115--325 \AA\ produce maximum O($^1$D)
atoms via photodissociation of CO. 
The dissociative excitation cross section of CO is maximum in the
wavelength region 535--955 \AA\  (cf. Fig.~\ref{o-csc}), which results in the peak production of 
O($^1$D) via photodissociation of CO at 1000 km. 
More details on the attenuation of solar flux in high water production rate comets are given  
in \cite{Bhardwaj03}.

\begin{figure}[h]
\begin{center}
\noindent\includegraphics[width=22pc,angle=0]{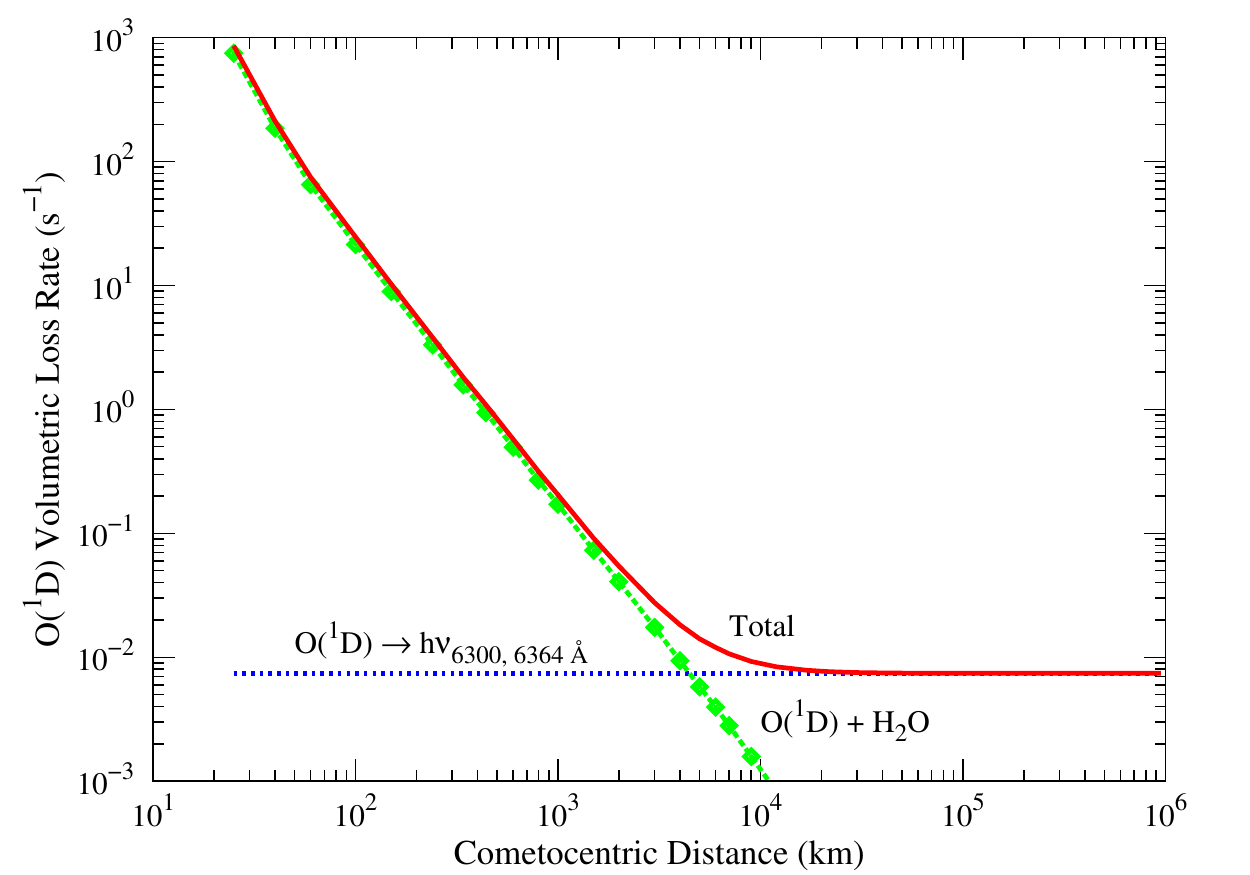}
\caption{Calculated radial profiles for major loss mechanisms of the O($^1$D) atom
 for the abundances of 6\%   CO$_2$ and 24\%   CO relative to H$_2$O production
rate of 8.3 $\times$ 10$^{30}$ s$^{-1}$.}
\label{lrate-o1d}
\end{center}
\end{figure}

The model calculated volumetric loss rate profiles of O($^1$D) are presented in Figure~\ref{lrate-o1d}. 
This figure depicts that the predominant destruction channel of O($^1$D) in the inner coma (below
3000 km) of comet \hbop\ is quenching by H$_2$O, which results in the formation of  two OH 
molecules. Above radial distance of 10$^4$ km,  the  radiative decay
leading to the red-doublet emissions is the major loss for O($^1$D) atoms. 
Quenching by CO$_2$ and CO are minor loss processes, about one order of magnitude smaller 
and hence is not shown. 

\begin{figure}[h]
\begin{center}
\noindent\includegraphics[width=22pc,angle=0]{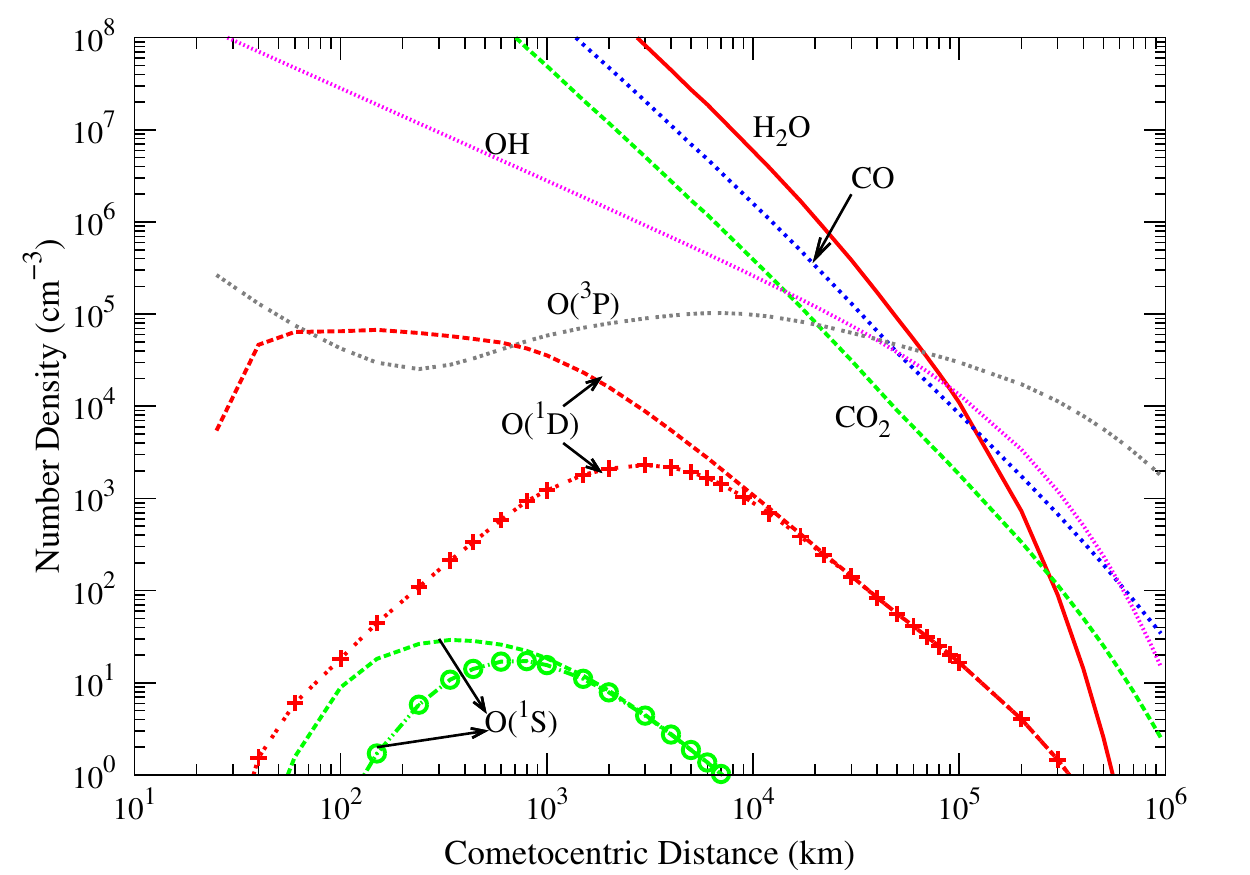}
\caption{Calculated number density profiles of O($^1$S), O($^1$D), O($^3$P), and OH, 
along with those of H$_2$O, CO, and CO$_2$. The calculations are done 
for the abundances of 6\%  CO$_2$ and 24\% CO relative to H$_2$O production
rate of 8.3 $\times$ 10$^{30}$ s$^{-1}$. The dashed lines of O($^1$S) and O($^1$D)
are the calculated densities  without accounting the collisional quenching processes.}
\label{den-osd}
\end{center}
\end{figure}

The calculated density profiles of O($^1$S), O($^1$D), and O($^3$P) in comet \hbop\ along with
 parent species considered in
our model  are shown in Figure~\ref{den-osd}. The  density of O($^1$S) peaks around 500 km, 
while the density profile of O($^1$D) shows a broad peak between 2000 and 5000 km.
The calculated number density profiles of O($^1$D) and  O($^1$S) without collisional quenching 
processes are also presented in this figure (with dashed lines). This calculation clearly shows 
that collisional quenching can significantly reduce the O($^1$S) and O($^1$D) densities   
in the inner coma. The formation of O($^3$P) below 200 km is due to collisions between OH molecules.

 \subsection{Forbidden emissions of atomic oxygen: [OI] 5577, 2972, 6300, and 6364 \AA}
\begin{figure}
\begin{center}
\noindent\includegraphics[width=22pc,angle=0]{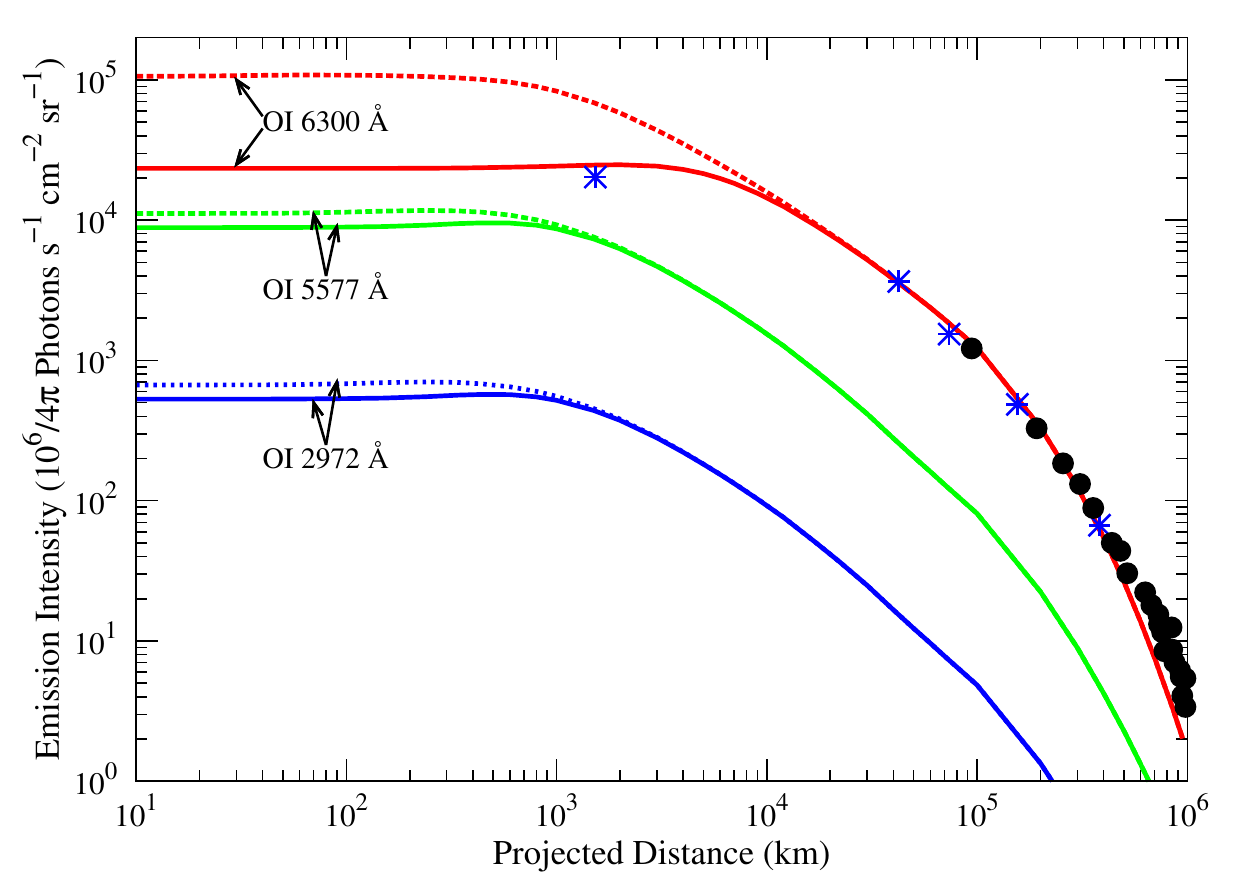}
\caption{Calculated [OI] 6300,  5577, and 2972 \AA\ line brightness profiles as a function of 
 the projected 
distances for the abundances of 6\%  CO$_2$ and 24\% CO relative to H$_2$O  production
rate of 8.3 $\times$ 10$^{30}$ s$^{-1}$. For comparison 
the observed intensities of 6300 \AA\ line emission by \cite{Morgenthaler01} using  Hydra (stars) 
and WHAM (filled circles) instruments on 1997 March 2 ($r_h$ = 1.05 AU and $\Delta$ = 1.46 AU) and 5 
($r_h$ = 1.03 AU and $\Delta$ = 1.42 AU), respectively, are also shown. During the observation 
the field of view of Hydra and WHAM instruments are 1$^\circ$ and 45$'$, respectively
 \citep{Morgenthaler01}. Dotted lines are the calculated intensities when collisional quenching is 
not accounted.} 
\label{proj-int}
\end{center}
\end{figure}

 The emission rates of [OI] 5577, 2972, 6300, and 6364 \AA\ are calculated by multiplying 
Einstein transition probabilities (A$_{5577}$ = 1.26 s$^{-1}$, A$_{2972}$ = 0.134 s$^{-1}$,
 A$_{6300}$ = 6.44 $\times$ 10$^{-3}$ s$^{-1}$,  and A$_{6364}$ = 2.17 
$\times$ 10$^{-3}$ s$^{-1}$) 
with the densities of O($^1$S) and O($^1$D) 
(see \cite{Bhardwaj12} for calculation details). The intensity of these line emissions along 
the line of sight is calculated 
by integrating the  emission rates. The model calculated brightness profiles as a 
function of projected distance  for these forbidden emissions along with the 
[OI] 6300 \AA\ observations  of
\cite{Morgenthaler01} made on 2 and 5 March  1997 using Hydra and WHAM instruments, respectively, 
 are presented in Figure~\ref{proj-int}. To show the collisional quenching effect, 
we also presented the calculated forbidden emission line intensities (with dotted lines) 
in Figure~\ref{proj-int},
 by  considering only radiative decay as the loss process of O($^1$S) and O($^1$D).
 The [OI] 2972 \AA\ emission profile is 
shown by taking  branching ratio of 5577/2972 as 10 as suggested by \cite{Slanger06}. The 
NIST recommended value for this ratio is 16 \citep{Wiese96}. 

 \renewcommand{\thefootnote}{\fnsymbol{footnote}}
\begin{table*}[tbh]              % Table 2
\begin{center}
 \small
\caption{Calculated percentage contributions for the major production processes of 
O($^1$S) and O($^1$D) in comet \hbop\ with varying relative abundance of 
 CO$_2$ for 0.5 \% O($^1$S) yield.} 
\scalebox{0.61}[1]{
\label{tabprj}
\begin{tabular}{ccccccccccccccccccccccccccccc}
\hline 
\multicolumn{1}{c}{CO$_2$} &  
\multicolumn{24}{c}{Production processes of O($^1$S) and O($^1$D) at three
 cometocentric projected distances (km) (\%)}    \\ 
\cline{2-25}
(\%)&\multicolumn{4}{c}{ h$\nu$ + H$_2$O  } 
&\multicolumn{4}{c}{h$\nu$ + OH} 
&\multicolumn{4}{c}{h$\nu$ + CO$_2$}
&\multicolumn{4}{c}{h$\nu$ + CO} 
&\multicolumn{4}{c}{O($^1$S) $\rightarrow$  O($^1$D)} 
&\multicolumn{4}{c}{e + H$_2$O$^+$} \\ 
\hline \smallskip
&10$^2$&10$^3$&10$^4$&10$^5$&10$^2$&10$^3$&10$^4$ &10$^5$ &10$^2$&10$^3$&10$^4$  
&10$^5$&10$^2$&10$^3$&10$^4$ &10$^5$&10$^2$&10$^3$&10$^4$ &10$^5$&10$^2$&10$^3$&10$^4$ &10$^5$\\  
% \cline{2-25}
6  & 25 (77)\footnotemark[2]  & 31  (76) &  24 (49) & 6 (7) & 0.5 (6) & 1 (8) & 4 (33)& 14 (75) 
& 58  (7) &  50 (4)& 40 (2) & 23(1) &  6  (0.5) & 7  (0.5) & 8  (1) & 5 (0.5)& (6) & (8) & (8) & (5)& 2  (2) 
& 3 (2) & 9  (8) & 7 (5)\\ 
3  & 33 (82)  & 42  (80) & 33  (49) & 8 (8)& 1 (7) & 1 (9) & 5  (34)& 18 (77)& 47 (4) & 36 (2)& 
 26 (1)  &15 (0.5) &  8 (0.5) & 9 (1) & 10  (1) & 7 (0.5)& (5) & (6) & (6) & (4)& 2 (2) & 3 (2) & 11 (8)& 11 (5)\\ 
1  &  49 (85)& 57 (82) & 42 (51) & 10 (8)& 1 (7)& 2 (9) &7 (35) & 24 (78) & 27 (1) & 17 (0.5) & 11 (0.5) 
&7 (0.5) & 12 (1) & 13 (1) & 13 (1) & 9 (0.5) & (4)& (5) & (5) & (3) & 4 (2) & 5 (3)& 15 (7) & 14 (5) \\ 
\hline
\end{tabular}
}
\footnotemark[2]{\small  The values in parenthesis are for the  O($^1$D).}
\end{center}
\end{table*}

The calculated percentage contributions of various processes involved in the production of metastable 
O($^1$S) and O($^1$D) at different projected distances are presented in Table~\ref{tabprj}. 
For 6\% relative abundance of CO$_2$,  
 photodissociation of CO$_2$ is the major source  of O($^1$S)  production rather than photodissociation
of H$_2$O (cf.~Fig.~\ref{prat-o1s}). 
 So we varied the CO$_2$ relative abundance  to study  the change in 
the contribution of CO$_2$ to the  O($^1$S) and O($^1$D) production. Calculations presented 
in Table~\ref{tabprj}
depict that, for a 6\% relative abundance of CO$_2$, below 10$^4$ km projected distances,
 around 25 to 30\% of O($^1$S) production is  via 
photodissociation of H$_2$O, while 40 to 60\% production is through photodissociation of CO$_2$.
Though the relative abundance of CO in comet  \hbop\ is high ($\sim$25\%), the photodissociation of CO 
could contribute a  maximum of 10\% to the  O($^1$S) production. The 
dissociative recombination of H$_2$O$^+$ and  CO$^+$ together can contribute  
 10\% to the production of O($^1$S), whereas  photodissociative excitation of OH is  a 
minor ($<$5\%) source. At 10$^5$ km projected distance, the photochemical reactions mentioned 
in Table~\ref{tabprj} all together contributing 60\% of O($^1$S) and remaining is contributed by 
dissociative recombination of O-bearing ions.
 When the abundance of CO$_2$ is reduced to 3\%, below 10$^4$ km projected distance,
 photodissociation of H$_2$O (35 to 40\%)  and  CO$_2$ (30 to 50\%) contribute almost equally to the  
production of O($^1$S).

The major production process of O($^1$D) is the photodissociation of  H$_2$O, whose contribution is 
 60 to 80\% below 10$^4$ km projected distance (cf.~Table~\ref{tabprj}).
 Around 10$^4$ km the photodissociation
of OH is also a significant production source of O($^1$D)  and contributes around 20\%; 
but, in the inner coma the contribution of this process is  small ($<$10\%). Radiative decay of O($^1$S)
 and electron recombination of H$_2$O$^+$ contribute less than  10\% each. At 10$^5$ km projected distance,
most (75\%) of O($^1$D) is produced by photodissociation of OH and remaining is contributed by other
 reactions. The change in the relative abundance of CO$_2$ by 
a factor of 2, from 6\% to 3\%, does not affect the relative contributions of various sources 
of O($^1$D) below 10$^4$ km projected distance.

\begin{center}
\begin{table}[h]
\small
 \caption{Calculated percentage contributions for the major production processes
 of green and red-doublet emissions in the total observed projected field of view 
(2.4 $\times$ 10$^{5}$ km) on comet \hale\ with varying relative abundance of CO$_2$.}
\scalebox{0.6}[1]{
\label{tab-slit}
\begin{tabular}{ccccccccccccccccccccccc} 
\hline
 \multicolumn{1}{c}{CO$_2$ (\%)} 
&\multicolumn{1}{c}{h$\nu$ + H$_2$O} 
&\multicolumn{1}{c}{h$\nu$ + OH}
&\multicolumn{1}{c}{h$\nu$ + CO$_2$}
&\multicolumn{1}{c}{e$^-$ + CO$_2^+$}
&\multicolumn{1}{c}{e$^-$ + H$_2$O$^+$}
&\multicolumn{1}{c}{O($^1$S) $\rightarrow$ O($^1$D)} 
&\multicolumn{1}{c} {\centering h$\nu$ + CO}  \\
\hline
 6  & 23 (48)\footnotemark[3] &  4 (35)& 41 (3) & 8  (0.5) & 7 (5) & (7)  & 7  (1)   \\  
 3  & 32 (50) &  6 (36)& 30  (2) & 5 (0.5) & 10 (8) & (7)  & 10 (1)      \\
1 & 42 (50) & 8 (37) & 13 (0.5) & 2 (0.5) & 13 (8) & (5)& 13 (1)\\
\hline
\end{tabular} 
}

\footnotemark[3]{The values in parenthesis are the calculated percentage  contributions for 
red-doublet emission.} 
\end{table}
\end{center}

For a 4$'$ circular aperture projected field of view ($\sim$2.4 $\times$ 10$^{5}$ km) on comet \hbop, 
which is similar to the 50 mm Fabry-P{\'e}rot 
spectrometer observations of \cite{Morgenthaler01},  the calculated  percentage contribution of 
major production 
processes for the green and red-doublet emissions, for different relative abundances of CO$_2$, 
are  presented in  Table~\ref{tab-slit}.
These calculations clearly suggest  that in a  comet which has been observed over a large projected 
area, the photodissociation of H$_2$O and OH mainly ($\sim$ 80\%) controls the [OI] 6300 \AA\ emission, 
while the radiative decay of O($^1$S) contributes a maximum value of 10\% to the total red-doublet intensity.
With 6\% relative abundance of CO$_2$, the [OI] 5577 \AA\ line emission  observed in the coma is
 largely ($\sim$40\%)
contributed by photodissociation of CO$_2$, and photodissociation of H$_2$O is the next significant
 ($\sim$25\%) 
production process. The other production processes, like dissociative recombination of ions, 
photodissociation of CO, OH, etc, together 
contribute less than 30\% to the [OI] 5577 \AA\ intensity. When the CO$_2$ abundance is reduced 
to 3\%, both photodissociation of H$_2$O and CO$_2$ are contributing equally ($\sim$30\%) to the 
 green line emission intensity. In all these  cases, in spite of CO relative abundance being
 high ($\sim$25\%) in comet \hbop, the photodissociation of CO could contribute a maximum value of 10\%. 

\subsection{Green to Red-doublet intensity ratio}
In comets, the parent species of these atomic oxygen emission lines are assessed using  
 the ratio of intensity of  the green line to the sum of
intensities of the red-doublet, which can calculated as
\begin{equation}
 \frac{I_{5577}}{I_{6300} + I_{6364}} = \frac{\tau^{-1}_{green} \alpha_{green}N_{green} \beta_{green}}
{\tau^{-1}_{red}\alpha_{red}N_{red}(\beta_{6300+6364})} 
\end{equation}
where $\tau$ is the lifetime of excited species in seconds ($\tau$[O($^1$D)] $\approx$ 110 s and 
$\tau$[O($^1$S)]  $\approx$ 0.7 s), $\alpha$ is the 
yield of photodissociation \citep{Huebner92}, $\beta$ is 
the branching ratio ($\beta_{6300}$ = 0.75, $\beta_{6364}$ = 0.25, $\beta_{5577}$ = 0.90, 
and $\beta_{2972}$ = 0.10
\citep{Wiese96,Slanger11,Festou81} of the transition, 
and N is the column density of cometary species in cm$^{-2}$.
Customarily, the observed G/R ratio of 0.1 has been used to 
confirm the  parent species of these oxygen lines as H$_2$O in comets \citep{Cochran84,Cochran08,
Morrison97,Zhang01,Cochran01,Furusho06,Capria05,Capria08,Capria10}. However, since no experimental 
cross section or yield for the production of O($^1$S) from H$_2$O is available
in the literature, this ratio has been questioned by \cite{Huestis06}. In our previous work 
\citep{Bhardwaj12}, by fitting the observed green line emission intensity in comet \hyak, 
 we suggested that the yield for  photo-production rate of O($^1$S) from H$_2$O
at solar H Ly-$\alpha$  can not be more than 1\%. Our previous model calculation also demonstrated that 
the determined G/R ratio depends on the projected  area observed over the comet.

\begin{figure}[h]
\begin{center}
\noindent\includegraphics[width=22pc,angle=0]{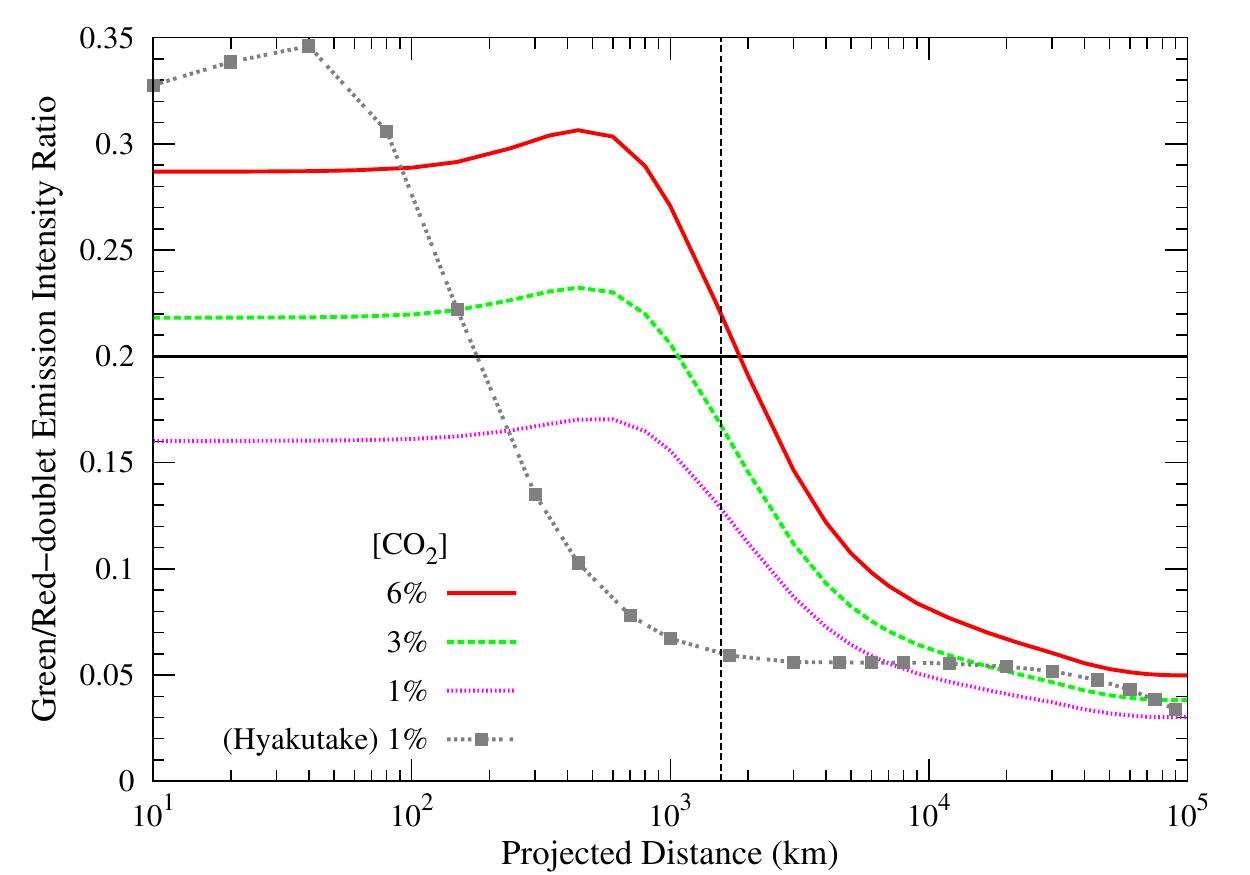}
\caption{\small Calculated green to red-doublet intensity ratio along projected distances 
for different CO$_2$ relative abundance [CO$_2$] and with 0.5\% yield  for O($^1$S) production 
 in the photodissociation of H$_2$O. \cite{Zhang01} observed average green to red-doublet 
intensity ratio was 0.2 for the slit projected size of 522 $\times$ 1566 km over comet \hbop\  on 
28 March 1997, which is shown with a horizontal line. The vertical dotted line represents 
1566 km projected distance on the cometary coma. For comparison the calculated G/R ratio 
profile with 1\% CO$_2$ and 0.5\% yield in comet \hyak\ is also shown.}
\label{gr-ratio}
\end{center}
\end{figure}
 
We calculated the G/R ratio profiles on comet \hbop\ on 26  March 1997  by varying CO$_2$ 
relative abundance from 6 to 3 to 1\% which are presented in Figure~\ref{gr-ratio}. 
 For comparison, the  G/R ratio profile calculated on comet \hya\ \citep{Bhardwaj12} is also plotted 
in Figure~\ref{gr-ratio}. In comet \hya\ the G/R ratio is constant up to 100 km projected distance, 
while in the case of comet \hbop\ it is constant even up to 1000 km. The flatness of the G/R ratio depends on 
the quenching rate of metastable O($^1$S) and O($^1$D) by H$_2$O which is a function of water 
production rate of the comet. Thus, in comets where H$_2$O production rate is still larger than that 
of \hbop, the G/R ratio would be constant up to projected distances larger than 10$^3$ km.

\subsection{Radiative efficiencies of O($^1$S) and O($^1$D) atoms}
\begin{center}
\begin{figure}[h]
\noindent\includegraphics[width=22pc,angle=0]{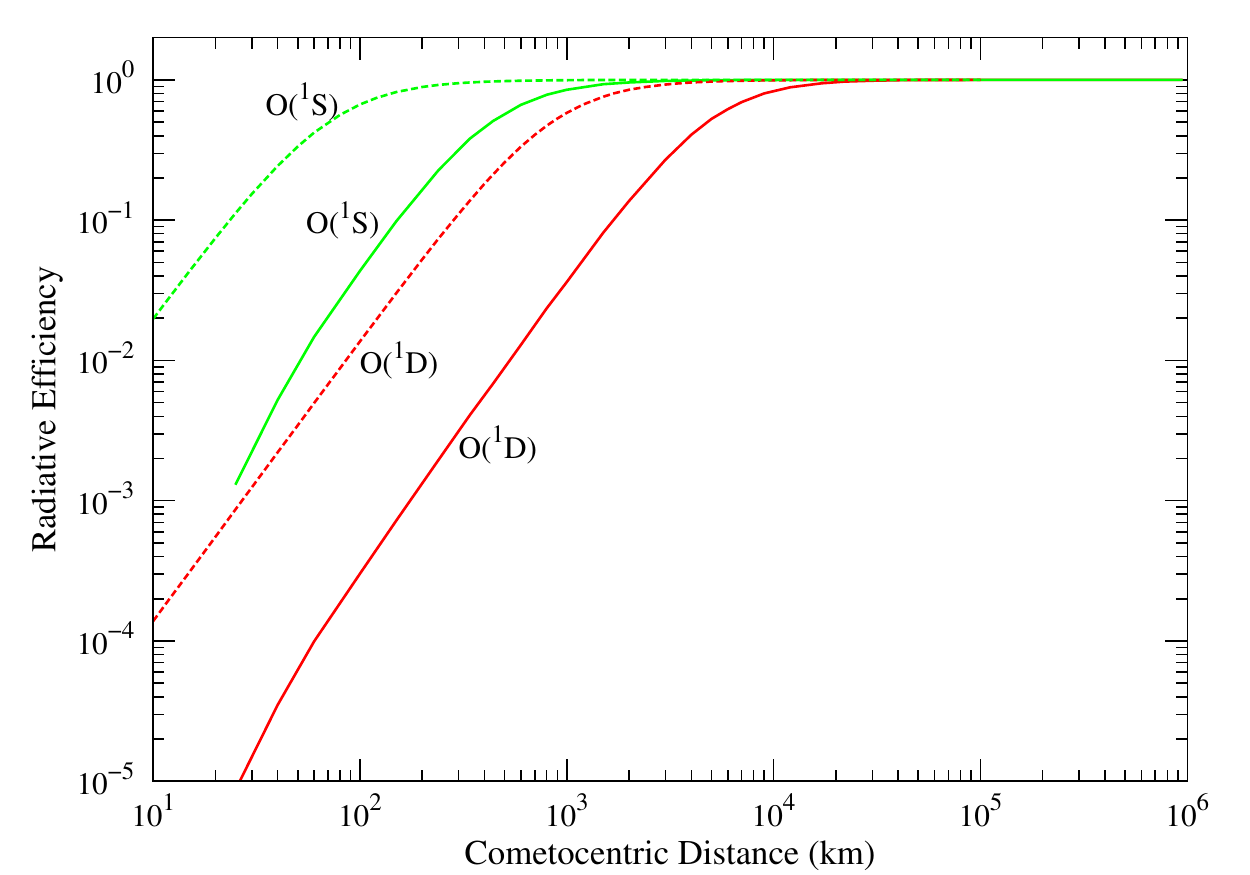}
\caption{The calculated radiative efficiency profiles of 
O($^1$S) and O($^1$D) on comets \hbop\ (solid lines) and  \hya\ (dashed lines). 
Radiative efficiency is the ratio of emission rate to the production rate.}
\label{radeff}
\end{figure}
\end{center}

The number density of O($^1$S) and O($^1$D) in the cometary coma is controlled by 
various production and loss processes at that radial distance.
To understand the region of maximum emission of green and red-doublet lines in the  coma 
 we calculated the radiative efficiency profiles of O($^1$S) and O($^1$D) in comets \hbop\ 
and \hya\ by calculating the ratio of emission rate to total production rate of
respective species. The calculated radiative efficiency profiles of O($^1$S) and O($^1$D) 
are presented in Figure~\ref{radeff} with 
solid and dotted line for comets \hbop\ and \hya, respectively. 
 This figure depicts that in comet \hbop\ all the O($^1$S) atoms produced 
above 1000 km radial distance emit 5577 \AA\ 
(or 2972 \AA) photons, while for O($^1$D) the radiative efficiency is unity 
 above 10$^4$ km. Since the lifetime of O($^1$D)
is higher by two orders of magnitude than that of O($^1$S), most of the produced O($^1$D)
in the inner coma get quenched by other cometary species (mainly by H$_2$O) without emitting photons 
at wavelengths 6300 and 6364 \AA. But in case of comet \hya\ the radiative efficiency of O($^1$S)
and O($^1$D) is unity above 100 and 1000 km, respectively. This calculation shows that in comets most of 
the green and red-doublet emissions are produced above the collisional-dominated region where the radiative 
decay is the dominant loss process for O($^1$S) and O($^1$D) atoms.

\subsection{Excess velocities  of O($^1$S) and O($^1$D)}
Solar photons having energy more than the dissociation threshold of cometary species impart the 
additional energy to the kinetic motions of daughter products.                 
The mean excess energy released in the $i$th dissociation  process at a radial	
 distance $r$  can be determined as
\begin{equation}
 E_i(r) =  \frac{\int_0^{\lambda_{th}} hc \left( {\frac{1}{\lambda}-\frac{1}{\lambda_{th}}}\right)
\sigma(\lambda)\phi(\lambda, r) e^{-\tau(\lambda, r)}d\lambda}
{\int_0^{\lambda_{th}} \sigma(\lambda)\phi(\lambda, r) e^{-\tau(\lambda, r)}d\lambda} 
\end{equation}
where $\lambda$ is the wavelength of solar photon,  $\lambda_{th}$ is the threshold wavelength 
for the dissociation process, $h$ is Planck's constant, and $c$ is the velocity of light. 
$\sigma(\lambda)$ is the dissociation cross section
 of the cometary species at wavelength $\lambda$. $\phi(\lambda, r)$ and $\tau(\lambda, r)$ are 
the solar flux and the optical depth of the medium for the photon of the wavelength $\lambda$ 
at a radial distance $r$, respectively.

\begin{center}
\begin{figure}[h]
\noindent\includegraphics[width=22pc,angle=0]{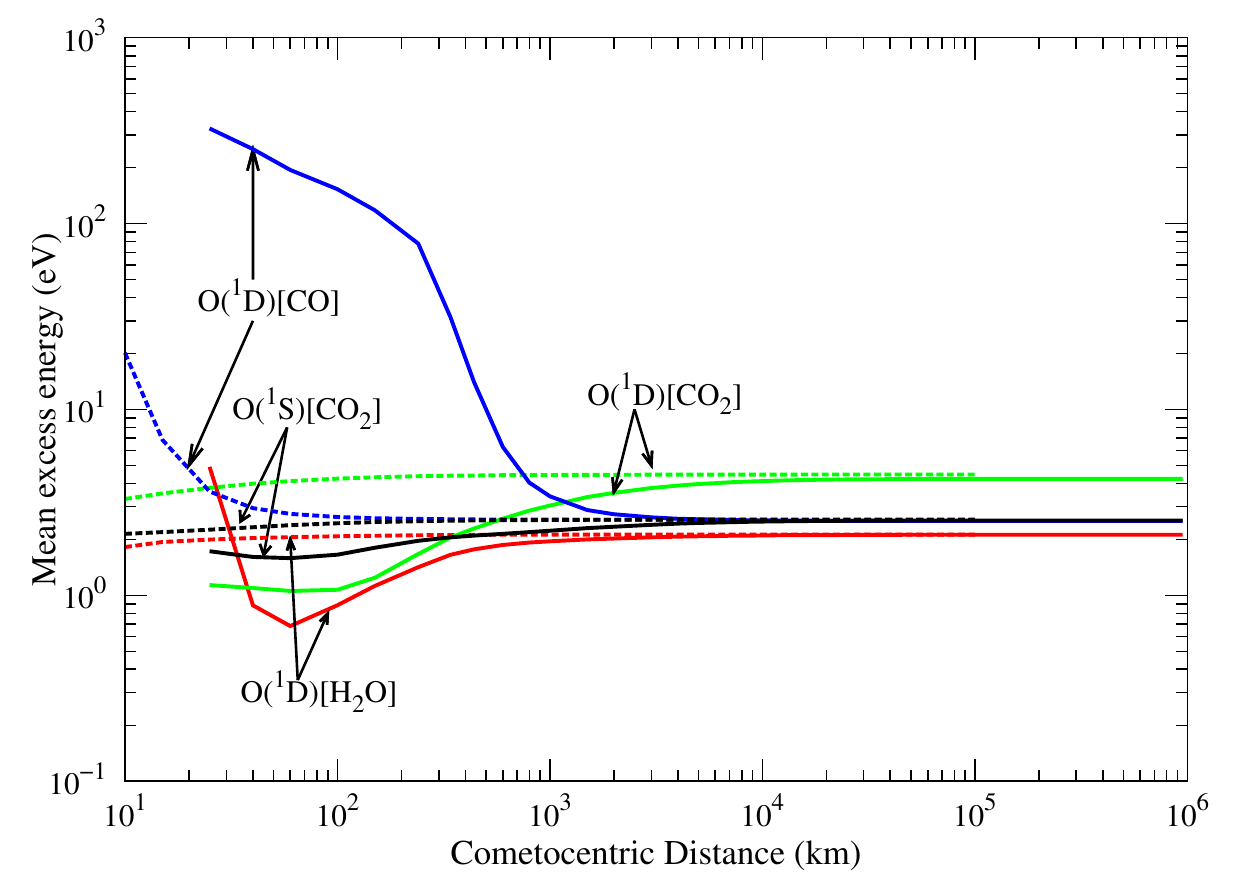}
\caption{Calculated excess energy profiles of  O($^1$D) in photodissociation 
of H$_2$O, CO, and CO$_2$ and that of O($^1$S) in photodissociation of CO$_2$
on comets \hbop\ (solid lines) and  \hya\ (dashed lines).}
\label{exvel}
\end{figure}
\end{center}

 Our model calculated 
mean excess energy profiles for the photodissociation of H$_2$O, CO$_2$, and CO  forming 
O($^1$S) and O($^1$D) are presented in Figure~\ref{exvel}  with
solid and dotted lines for comets Hale-Bopp and Hyakutake, respectively. 
 Above 3000 km radial distance,  the calculated excess energies 
in different photodissociation processes in both comets show a constant profile, because 
the optical depth in this region for photons of different wavelengths is very small. These values are 
in agreement with the calculations of \cite{Huebner92}.
 However, at shorter radial distances the neutral density is higher and hence 
the wavelength dependent photodissociation is significant which causes different excess energy values.

In comet \hbop\ the calculated mean excess energy in photodissociation of H$_2$O producing O($^1$D)
shows a highest value of 5.6 eV at the surface of the nucleus and decreases to a minimum value 
 of 0.7 eV at 50 km. Above 50 km the mean excess energy   
increases  and becomes  constant (2.12 eV) above  3000 km.
This is because of the formation of O($^1$D) via the photodissociation of H$_2$O
 is associated with the photons of different energies and it also varies with 
radial distance as shown in Figure~\ref{wav-o1d-h2o}.
 At a given radial distance the mean excess energy released in the photodissociation process
  is determined by the mean of energies of different solar photons involved. 
The threshold energy for production of O($^1$D) by dissociating H$_2$O  is 7 eV.
 Very close to the cometary nucleus
 ($<$50 km), photons of wavelength  smaller than 115 \AA\ and in the wavelength band 1375--1575 \AA\ 
determines the formation of O($^1$D) (cf.~Fig.~\ref{wav-o1d-h2o}). 
At this distance, most of O($^1$D) is  produced by the photons of low energy (7--9 eV)
in the wavelength band 1375--1575 \AA, and a small amount of O($^1$D) is produced by 
very high energy ($>$100 eV) photons which results in the mean excess energy of about 2--5 eV. 
But around 50 km,  the majority of O($^1$D)  production 
  is determined by the photons of low energy 7 to 12 eV 
(955--1575 \AA\  wavelength band) and
the contribution from photons of wavelength  below 115 \AA\ is very small. This causes 
the minimum value of mean excess energy 0.7 eV  at this radial distance.

Between 50 and 300 km, the increase in the excess energy is due to the production of
O($^1$D) atoms by photons of wavelength bands
115--325, 955--1575 \AA, and  solar H Ly-$\alpha$. 
Though high energy photons (115--325 \AA) are also involved 
 in this region, the intense solar photon flux at H Ly-$\alpha$ (1216 \AA)  governs
 the majority of O($^1$D) production and subsequently determines the mean excess energy. 
The solar H Ly-$\alpha$ photons can provide the maximum excess energy of 3 eV 
in the photodissociation of H$_2$O. 
Above 1000 km more than 90\% of the O($^1$D) production is controlled by 
photons at 1216 \AA\ wavelength and the remaining from other wavelength bands 
(cf.~Fig.~\ref{wav-o1d-h2o}), which results a constant value of mean excess energy of 2.12 eV.

Similarly, the mean excess energy  calculated in the photodissociation of CO$_2$ producing
O($^1$D) can be explained based on the wavelength dependent photon attenuated profiles presented
 in Figure~\ref{wav-o1d-co2}. The threshold energy for the O($^1$D) production in 
photodissociation of CO$_2$ 
   is 7 eV and for O($^1$S) it is 9 eV. At radial distances less than 100 km, the production of O($^1$D) in
photodissociation of CO$_2$ is determined by the photons of low energy (average 8 eV)
 in the wavelength 
bands  1375-1785 \AA\ and 955-1165 \AA, which results in low mean excess energy of $\sim$1 eV. Above 100 km,
photons of different energies ranging from 7 to 16 eV (cf.~Fig.~\ref{wav-o1d-co2}) causes the mean 
excess energy of $\sim$4 eV.
The calculated mean excess energy profiles  in the photodissociation of CO$_2$ producing O($^1$S) and 
O($^1$D) are not similar. This is because the O($^1$S) production 
 occurs via photodissociation of CO$_2$ in the wavelength band of 800 to 1300 \AA\ 
(photons of 10-15 eV),
whereas O($^1$D) can be produced by photons of wavelength less than 800 \AA\ 
($>$15 eV) (cf. Fig.~\ref{o-csc}). 

The threshold energy for the dissociation of CO producing O($^1$D) is 14.3 eV. 
Below 200 km the calculated maximum mean excess energy in the  photodissociation of CO producing 
O($^1$D)  is more than 100 eV. This is because the formation of O($^1$D) at these distances
(cf.~Fig.~\ref{wav-o1d-co}) is mainly
determined by photons of wavelength less than 115 \AA\ ($>$110 eV) with some contribution from
 the wavelength  band  115--325 \AA\ (40--110 eV). Above 500 km, the formation of O($^1$D)
 is mainly due to solar photons  in 
the wavelength band 535--955 \AA\ (23--13 eV) which results in the maximum excess energy of 2.5 eV.

 \section{Discussion}
The major difference between comets \hbop\ and \hya\ is the H$_2$O production rate, which is
larger by a factor of  30 in  the former. This difference in the H$_2$O production rates 
result in a change
 in  the photochemistry of O($^1$S) and O($^1$D) in the cometary coma. Due to the 
dense coma of comet \hbop, the attenuation of solar UV-EUV 
photons on \hbop\  differs significantly from that in \hya. 
Moreover, the CO$_2$ abundance in comet \hya\ is smaller ($<$3\% relative abundance) compared to  
that in \hbop\ ($\sim$6\% relative abundance). The high H$_2$O production rate in comet
 \hbop\ results in
a larger collisional coma (radius few $\times$ 10$^5$ km) which is comparable to the
scale length ($\sim$8 $\times$ 10$^4$ km)  of H$_2$O molecule. In the low production rate 
comets the collisional zone is smaller and  photochemistry significantly  differs.

The photodissociation rates of H$_2$O and CO$_2$ for O($^1$S)
 production differ by a factor of 20 (cf. Table~\ref{tabo1sd}). Hence, 
 the major source of O($^1$S) in the inner  coma of comet \hbop\ is 
  photodissociation of CO$_2$ rather than photodissociation of H$_2$O.
Since the relative abundance of 
CO$_2$ in comet \hya\ is 1\%, the photodissociation of CO$_2$ becomes an important
source only near the surface of the nucleus \citep[cf.~Figure 6~of][]{Bhardwaj12}. 
The  production peak of O($^1$S) in comet \hya\
is closer to the nucleus ($<$20 km), whereas in comet \hbop\ it is between 100 and 1000 km. 
Even when we reduced the CO$_2$ abundance by 50\% in \hbop, the peak production of O($^1$S) 
in the inner coma is mainly controlled  by photodissociation of CO$_2$ 
and not by photodissociation of H$_2$O. 
Hence, in a high water production rate comet a small relative abundance ($\sim$5\%) of CO$_2$,
makes CO$_2$ as a potentially important source of O($^1$S) compared to H$_2$O.

 In comet \hya, inside 10$^5$ km,  the photodissociation of H$_2$O is the major 
(more than 90\%) production process  
 of O($^1$D) formation and the contributions from other processes are very small.
But in comet \hbop, since the H$_2$O production rate and CO$_2$ relative abundance are higher,
 the solar photons of wavelength 955--1165 \AA, which are less attenuated by H$_2$O, 
can travel deeper into the cometary coma and dissociate the CO$_2$ to form O($^1$D),
 which is not the case in  comet \hya.

The radius of collisional coma, which is a function of total gas production rate, 
 in comets \hya\ and \hbop\ differs by an order of magnitude.
 In comet \hya\  quenching of O($^1$S) by 
 H$_2$O is the main destruction mechanism only close to the nucleus ($<$50 km)
and radiative decay dominates at distances larger than 100 km. However, in comet
\hbop\ collisional quenching is significant up to 500 km and only above that 
radiative decay is the major loss mechanism of O($^1$S).
Similarly, the collisional quenching radii of O($^1$D) in comets \hya\ ($\sim$10$^3$ km) 
and \hbop\ ($\sim$10$^4$ km)  also differs by an order of magnitude.

The  O($^1$D) density peak in 
comet \hbop\ is broader (2000 to 5000 km) than that in comet \hya\ (200 to 600 km). 
This change in the peak distribution of O($^1$D) in the two comets is due to 
different H$_2$O production rates and  wavelength dependent photo-attenuation 
in the cometary comae.

\subsection{Comparison of model calculations with observations}
\subsubsection{[OI] 6300 \AA\ emission}
\cite{Morgenthaler01}  observed [OI] 6300 \AA\ emission on comet \hbop\ on several days 
during February to April 1997 using 4 different ground based instruments. 
Large aperture observations of 6300 \AA\ emission using WHAM and Hydra spectrometers are made 
for the field of view  1$^\circ$ and 45$'$, which covers projected distances of 
1.5 $\times$ 10$^6$ and 2.4 $\times$ 10$^5$ km  on the comet, respectively. Our model calculated 
brightness profile of 
[OI] 6300 \AA\ emission shown in Figure~\ref{proj-int} is consistent with these observations.
 The brightness profile of [OI] 5577 \AA\ starts falling off beyond 1000 km, while the [OI] 6300 \AA\ 
profile remains constant up to 5000 km. The flatness in the calculated surface brightness profiles 
mainly depends on the collisional quenching of metastable species which is a function of H$_2$O 
production rate. The calculated green and red-doublet emission intensities (dotted lines in
Fig.~\ref{proj-int}) 
when radiative decay is considered as the only loss mechanism shows the role of 
collisional quenching. Since the lifetime is relatively larger, the O($^1$D) is substantially 
quenched by H$_2$O in the inner coma. Hence, below 1000 km, the calculated [OI] 6300 \AA\ emission
intensities differ by a factor of 5.

\begin{table}[t]
%   \begin{center}
\caption{The model calculated intensities of forbidden atomic oxygen emission lines 
on comet \hbop\ and the comparison of [OI] 6300 \AA\ line with the observation 
of \cite{Morgenthaler01} with 3\% CO$_2$ and 24\% CO.}
\label{tab-day}
\scalebox{0.65}[0.8]{
\begin{tabular}{lccccccccc}
\hline
\multicolumn{1}{p{1.5 cm}}{Date on} & \multicolumn{1}{c}{r} &
\multicolumn{1}{c}{$\Delta$} & 
\multicolumn{5}{c}{Intensity (R)} \\ %[-15pt]
\cline{4-8}
 1997 &  (AU)  & (AU)  & 2972 \AA & 5577 \AA & 6364 \AA & \multicolumn{2}{c}{6300 \AA} \\
\hline
 &    & &  & & &  Calculated\footnotemark[3] & Observed\footnotemark[4]  \\
% \cline{7-8}
Mar 9  & 0.999 & 1.383 & 34 & 330  & 1162 & 3637  & 2580--2922   \\
Mar 10 & 0.992 & 1.373 & 36 & 339  & 1192 & 3730  & 2300--2649   \\
Apr 7  & 0.920 & 1.408 & 45 & 423  & 1422 & 4450  & 2915--4964   \\
Apr 8  & 0.923 & 1.420 & 43 & 416  & 1400 & 4379  & 3057--3496   \\
Apr 9  & 0.925 & 1.431 & 43 & 411  & 1380 & 4323  & 2920--3197 \\
Apr 10 & 0.928 & 1.444 & 43 & 403  & 1358 & 4248  & 1579--1669 \\
Apr 13 & 0.939 & 1.484 & 39 & 372  & 1271 & 3296  & 1451--1960 \\
Apr 14 & 0.943 & 1.497 & 37 & 361  & 1240 & 3878  & 1575--2360 \\
Apr 16 & 0.952 & 1.526 & 36 & 339  & 1179 & 3688  & 2335--2974 \\
\hline
\end{tabular}}

\footnotemark[3]{\small The calculated average surface brightness over the observed
 projected distance of 2.5 $\times$ 10$^5$ km.}

\footnotemark[4]{\small The upper and lower limits of [OI] 6300 \AA\ intensity 
observed  by \cite{Morgenthaler01}.}
%   \end{center}
\end{table}

We also calculated the [OI] 6300 \AA\ emission intensity for  a circular aperture of 4$'$ diameter 
 on different days of March and April 1997 
 similar to the observation conditions of \cite{Morgenthaler01}.
 The calculated  intensities 
of different atomic oxygen emissions are presented in Table~\ref{tab-day} along with the
 [OI] 6300 \AA\ emission
intensities observed by \cite{Morgenthaler01}. Our calculated intensities are higher by a factor 
of 1.5 to 2.5 compared to the observation and also vary for different days  due to change in  solar flux,
 H$_2$O production rate, and heliocentric and geocentric distances. 
The observed [OI] 6300 \AA\ intensity on comet \hbop\ on 7 April 1997 is found to vary by a
factor of 1.6 in a span of less than 20 minutes, which is difficult to explain with the variation of 
heliocentric dependent water production rate. Similarly, the observed intensity values 
on the other days of observation also show large variation. The brightness during April 10 to 14
is consistently lower than during April 7 to 9. The variation in the observed intensity might be
 associated with spectral extraction process because of non uniform sensitivity of Fabry-P{\'e}rot 
spectrometer \citep{Morgenthaler01}, rather than the intrinsic variation in the comet.

\subsubsection{Green to red-doublet intensity ratio}
\cite{Zhang01} observed comet \hbop\ on 26  March 1997 using a rectangular slit  
(1.06$''$ $\times$ 3.18$''$) when the comet was 
at a geocentric distance of 1.32 AU and  heliocentric distance of 0.92 AU. For this observation,
the projected field of view on the comet  was 522 $\times$ 1566 km.
Our calculated  G/R  ratio with 3\% relative abundance of CO$_2$ and 
0.5\% yield of O($^1$S)  is 0.21, which is consistent with the observed G/R ratio
range (0.18--0.22) of \cite{Zhang01}. 
The calculated average G/R ratio, for a 4$'$ circular aperture field of view
with 3\% relative abundance of CO$_2$  for the different days of observation 
presented in  Table~\ref{tab-day}, is around 0.1. This shows that in a 
high water production rate comet the observed G/R ratio over a large projected distances 
($\sim$10$^4$ km) can be around 0.1 (cf.~Fig.~\ref{gr-ratio}). 
However, the calculated contributions of different 
production processes for O($^1$S)  suggest that photodissociation 
of CO$_2$ is more important  source rather than the photodissociation of H$_2$O. 
Hence, in comets with sufficient CO$_2$ abundances ($\ge$5\%), the green line emission is 
largely controlled by photodissociation of CO$_2$
and the derived G/R ratio over large cometocentric distances could be around 0.1. 
  \begin{center}
\begin{table}[tbh]
\small
\caption{The model calculated green and red-doublet emission intensities and the derived
O($^1$D) and H$_2$O production rates for different slit dimensions. 
The calculations are done with Q(H$_2$O) = 8.3 $\times$ 10$^{30}$ s$^{-1}$
for the relative abundances of 6\% CO$_2$ and 24\% CO at r$_h$ = 1 AU and $\Delta$ = 1 AU using
solar flux on 10 April 1997 (solar minimum period : solar radio flux 
F10.7 = 74.7 $\times$ 10$^{-22}$ J s$^{-1}$ m$^{-2}$ Hz$^{-1}$).}
\label{tab-slt}
\scalebox{0.67}[0.8]{
\begin{tabular}{cccccccccc}
\hline
Slit dimension & \multicolumn{2}{c}{Average intensity (R)\footnotemark[2] } 
& \multicolumn{2}{c}{Production rate (s$^{-1}$)}  & G/R\footnotemark[7]\\[0.5pt]
 \cline{2-3}  \cline{4-5} 
 (Projected distance in km) &[OI] 6300 \AA& [OI] 5577 \AA& Q[O($^1$D)] & Q[H$_2$O]\footnotemark[3]&\\[0.5pt]
 \hline
2$''$ $\times$ 2$''$  (725) & 18895 [83188]\footnotemark[1] & 7584 [9245] & 1.3 $\times$ 10$^{26}$ 
 & 3.7 $\times$ 10$^{26}$& 0.30 (0.08)\footnotemark[8]\\ % 
5$''$ $\times$ 5$''$  (1.8 $\times$ 10$^3$)& 18909 [68301] & 6723 [7584] & 8.1  $\times$ 10$^{26}$ 
& 23 $\times$ 10$^{26}$& 0.21 (0.08)\\ %
10$''$ $\times$ 10$''$ (3.6 $\times$ 10$^3$)& 19021 [52977] & 5369 [5825] & 3.3  $\times$ 10$^{27}$ 
& 9.3 $\times$ 10$^{27}$& 0.21 (0.08)\\ % 
30$''$ $\times$ 30$''$ (1.1 $\times$ 10$^4$) & 15668 [29341] & 2963 [3118]& 2.5  $\times$ 10$^{28}$ 
& 7.1 $\times$ 10$^{28}$&0.14 (0.08)\\ % 
1$'$ $\times$ 1$'$   (2.2 $\times$ 10$^4$) & 11785 [18793] & 1846 [1924]& 7.6  $\times$ 10$^{28}$ 
& 2.1  $\times$ 10$^{29}$&0.10 (0.08)\\
4$'$ $\times$ 4$'$   (8.7 $\times$ 10$^4$) & 5005 [6767]& 605 [624]     & 5.0  $\times$ 10$^{29}$
& 1.4  $\times$ 10$^{30}$ & 0.09 (0.07)\\
10$'$ $\times$ 10$'$  (2.1 $\times$ 10$^5$) & 2351 [3056]& 263 [271]     & 1.4  $\times$ 10$^{30}$
& 3.9  $\times$ 10$^{30}$&0.08 (0.07)\\
\hline
\end{tabular}}

\footnotemark[1]{The values in the square brackets are the calculated intensities without 
accounting for collisional quenching of O($^1$S) and O($^1$D).}
\footnotemark[2]{\small Intensity is averaged over the projected field of view,
1 R = $\frac{10^6} {4\pi}$ Photons s$^{-1}$ cm$^{-2}$ sr$^{-1}$;}
\footnotemark[3]{\small The branching ratio for the production of O($^1$D) in the 
photodissociation of OH
is taken as 0.357 \citep[see][]{Morgenthaler01}, while for the photodissociation 
of H$_2$O producing O($^1$D) it is 0.064 (This work). The branching ratio (0.81)
 for the production of OH in photodissociation of H$_2$O is taken from \cite{Huebner92}.}
\footnotemark[7]{\small Green to red-doublet emission intensity ratio determined over 
the projected field of view.}
\footnotemark[8]{\small The calculated G/R ratio without collisional quenching.}
\end{table}
  \end{center}

To evaluate the role of slit dimension in determining the G/R ratio 
we calculated green and red line intensities for various slit sizes by 
keeping H$_2$O, CO and CO$_2$ production rates as a constant.
  These calculations 
are presented in Table~\ref{tab-slt}. By varying the slit dimension from 2$''$ $\times$ 2$''$ 
to 10$'$ $\times$ 10$'$ the calculated G/R ratio over the projected cometary coma 
 changed from 0.3 to 0.08.  This result
clearly shows that the G/R ratio depends 
 not only on the photo-chemistry in the coma but also on the projected area observed 
 for the comet. The calculated G/R ratio is 
a constant value (0.08) throughout  the cometary coma when collisional quenching is neglected 
in the model. By doubling the CO$_2$ relative abundance 
in the coma, the G/R ratio increases by 30\% whereas the collisional quenching
 of O($^1$D) and O($^1$S) can change its value even by an order of magnitude. 

Besides the dimension of the slit used for observation, the projected area observed on 
the comet depends on  geocentric distance of the comet. Hence in a comet, where the collisional 
coma is resolvable in the observation, the derived G/R ratio depends on the projected area and also
on the collisional quenching of O($^1$S) and O($^1$D) in the cometary coma. 
Thus, we conclude that the observed G/R ratio of 0.1  is not 
a definitive benchmark value to verify  H$_2$O or CO$_2$/CO as the parent sources of atomic 
oxygen visible emissions in comets.

\subsubsection{Width of green and red-doublet emission lines}
 \cite{Cochran08} has found that the width of green line is higher than either of the  
red-doublet lines in the spectra of 8 comets. The wider green line implies the higher mean velocity  
of metastable O($^1$S), which could be  associated with different production processes. 
Besides collisions with different cometary species, the mean velocity 
of O($^1$S) in the cometary coma is determined by various production processes,
and/or could be due to the involvement of  photons of various energies in dissociating
O-bearing species \citep{Cochran08}.

The observed width of forbidden line emission depends on the velocity distribution of 
radiating metastable oxygen atoms. We found that the excess velocity 
released in photodissociation H$_2$O in the unity radiative efficiency region is 2.1 eV 
(cf. Figure~\ref{exvel}). If we assume that most of this excess energy is 
transfered to kinetic motion of atomic oxygen then the 
maximum mean velocity that can be acquired by the O($^1$D) atom would be 1.6 km s$^{-1}$. 
This velocity is consistent with values of 0.5 to 1.8 km s$^{-1}$ derived by 
\cite{Cochran08} in 8 comets. This supports the idea that most of the
red-doublet emission in cometary coma is governed by the photodissociation 
of H$_2$O. The excess energy profiles shown in Figure~\ref{exvel} suggest 
that the O($^1$D) produced in photodissociation of CO and CO$_2$ will have higher velocity
 than that produced in photodissociation of H$_2$O. The excess energy released in the 
photodissociation of CO and CO$_2$ in the unity radiative efficiency region is 2.5 eV and 4.1 eV,
 which corresponds to O($^1$D) excess velocity of $\sim$3.7 km s$^{-1}$ and 4 km s$^{-1}$, respectively.
However, our calculations suggest that CO and CO$_2$ together can contribute to a 
maximum  of 10\% to the red-doublet emission. 
The contributions of CO and CO$_2$ in the wings of red-doublet lines are probable. 

In the case of green line emission, since there is no experimentally determined cross section or yield
 for the photodissociation of H$_2$O 
 producing O($^1$S), it is difficult to determine the mean velocity acquired by an O($^1$S) atom 
in the the photolysis of H$_2$O. The maximum excess energy that can be released in photolysis of H$_2$O 
producing O($^1$S) 
at solar H Ly-$\alpha$ is 1.27 eV. Again, if we assume all the excess energy is transferred
as kinetic energy of atomic 
oxygen in $^1$S state then the maximum excess velocity of O($^1$S) would be 1.3 km s$^{-1}$. 
But in the case of photodissociation of CO$_2$, the excess energy is 2.5 eV, which corresponds to a 
 maximum O($^1$S) velocity of 4.3 km s$^{-1}$.
 The dissociative recombination of 
ions H$_2$O$^+$, CO$_2^+$, and CO$^+$ can contribute a maximum of 30\% in the production of 
green line emission. 
But the excess energy released in these recombination reactions is very small \citep{Rosen00,Rosen98,
Seiersen03}.  By assuming that the maximum 
mean velocity that can be acquired by O($^1$S) via the dissociative recombination processes is 
about 1 km s$^{-1}$, we found that the mean velocity of O($^1$S) from  all production processes 
is $\sim$2 km s$^{-1}$. This value is consistent with the derived velocity
 range of 1.9 to 3.1 km s$^{-1}$  for O($^1$S) in 8 comets by \cite{Cochran08}. 

Before coming
to a broad conclusion, we suggest that one has to calculate the exact mean excess velocities of 
O($^1$S) and O($^1$D) over the observed cometary coma, by accounting for all collisional 
processes and the mean excess velocity profiles of various species. Due to non availability 
of photon cross sections for some of the 
 photodissociation processes, and uncertainties involved in the excess energy calculations for
dissociative recombination reactions, 
our model is limited in determining the exact line widths of green and red-doublet emissions. 
However, based on our model calculations on comets \hbop\ and \hya, we 
suggest that involvement of multiple sources in the formation 
O($^1$S) could be  a potential reason for the higher line width of green emission compared to  that of  
red-doublet emission observed in several comets.

\subsection{Effect of model parameters on the calculated intensities}

\subsubsection{Expansion velocity of neutrals}
\label{eff-vel}
As we mentioned earlier in the Section~\ref{sec:model}, we have used the velocity profile 
from the work of \cite{Combi99} for calculating the number densities of parent species H$_2$O,
CO$_2$, and CO. \cite{Combi99} have shown that there is an acceleration of neutrals in
 the inner coma due to the photolytic heating \citep{Combi99,Colom99,Biver97,Combi02} 
and  other processes \citep{Harris02}. To evaluate the impact of this acceleration 
on our model results we carried out calculations by taking a constant gas expansion
 velocity profile with the values 0.7 and 2.2 km s$^{-1}$.
By using a constant velocity profile of 0.7 km s$^{-1}$ in the coma, rather than 
a radially varying velocity of \cite{Combi99}, the calculated intensities of green  
and red-doublet emissions are increased by 30\% and 25\%, respectively, which are 
still higher than the observation.
 By changing the constant gas expansion velocity from 0.7 to 2.2  km s$^{-1}$, the calculated 
intensities of atomic oxygen emission lines are decreased by $\sim$50\%. 
However using the \cite{Combi99}  velocity profile, our calculated [OI] 6300 \AA\ emission 
intensities over 4$'$ circular aperture field of view are closer to the observation 
(cf.~Table~\ref{tab-day}). 
 Hence, the velocity profile of neutral species is an important input in the model that 
should be accounted in calculating the intensities of these forbidden emissions. 

\subsubsection{Relative abundances of neutral species}
\label{relabn}
The water production rate in comet \hbop\ has been derived using emissions of direct and 
daughter products of H$_2$O by different observers \citep{Weaver97,Colom99,Schleicher97,Combi00,
Russo00,Woods00,Morgenthaler01,Harris02,Fink09}. During the observation period of these green and 
red-doublet emissions (r$_h$ of the comet was around 0.9 AU), \cite{Russo00} measured the 
H$_2$O production rates using  infrared
 emissions of water molecules
for different days. In this period, \cite{Combi00} derived the H$_2$O production rate 
 in this comet using H Ly-$\alpha$ emission. The difference between these two derived 
production rates is less than 
20\%. These observations found that around 1 AU the water production rate in comet \hbop\
was about $\sim$1 $\times$ 10$^{31}$ s$^{-1}$. Similarly, the derived water production rates of 
\cite{Fink09} on 1997 March 3 was 6.1 $\times$ 10$^{30}$ s$^{-1}$ which is smaller 
than the \cite{Combi00} derived rate by a factor of 1.5.
Using visible emission of atomic oxygen 
\cite{Morgenthaler01}  derived the H$_2$O production rates by applying standard branching ratios 
of OH and H$_2$O. These derived H$_2$O production rates are higher by factor of 3 to 6 compared to values
 determined from other observations. To assess the impact of H$_2$O production rate on 
the calculated green and red-doublet 
emissions we increased its value by a factor of 5. With increase in H$_2$O production rate 
the model calculated surface brightness of green and red-doublet emissions over 4$'$ circular 
field of view  is increased by a factor of 3. 

As demonstrated earlier in this paper, the role of CO$_2$ is very significant in determining 
the green line emission intensity and subsequently the G/R ratio. During the observation period
 of these forbidden emission lines the
CO$_2$ is not observed in this comet. To evaluate the impact of CO$_2$ we varied its relative 
abundance from 3 to 6\%. We found an increase (25\%) in the calculated green 
line emission intensity over the 4$'$ 
circular aperture field of view whereas it is small ($<$5\%) for red-doublet emission intensity.

Based on  infrared observations made  near perihelion on  
comet \hbop, \cite{Disanti01} suggested that 50\% of CO abundance present in the cometary 
coma is contributed by distributed sources. 
\cite{Bockelee10} investigated the extended distribution of CO by probing
  \hbop\ between $\sim$800 to $\sim$20,000 km region using
 CO rotational line emissions (viz, CO J(1-0)  and CO J(2-1)).
Based on the observation and radiative transfer modelling studies, 
\cite{Bockelee10} rejected the idea of an extended distribution of CO in \hbop. 
 Since the contribution of photodissociation of CO  to formation of O($^1$S) and O($^1$D)
is less than 10\%, no significant variation 
in the calculated intensity of green and red-doublet emissions is found by reducing the 
CO relative abundance by half. Hence, the involvement of CO in these oxygen forbidden 
line emissions is almost insignificant.

Though OH column densities are determined using 3080 \AA\ surface brightness profile,
there are large uncertainties in photo-cross sections of OH in 
producing O($^1$D) and 
O($^1$S) \citep{Huebner92,Morgenthaler01}.  
The calculated photo-rates for the production of O($^1$D) via photodissociation of OH,
  using theoretical and experimental cross sections differ by about an order of 
magnitude \citep{Huebner92}. \cite{Morgenthaler01} studied the effect of these cross 
sections in deriving the H$_2$O production rates using 
6300 \AA\ surface brightness profile and found that on
using the theoretical OH photodissociative branching ratios of O($^1$D),
the  derived H$_2$O production rates are higher by a factor of 3--6, than 
those determined based on experimental branching ratios of \cite{Nee84}.
  The photodissociation of OH influences the 
 calculated green and red-doublet emission intensities significantly above 10$^4$ km 
(cf. Figs.~\ref{prat-o1s} and \ref{prat-o1d}, and Table~\ref{tabprj}). 
By changing photorates determined by \cite{Nee84} experimental 
cross sections (which are used in the model) with the rates derived based on theoretically calculated
cross sections of \cite{Dishoeck84}, we found a 40\% decrease in the 
calculated slit-averaged brightness over the 4$'$ circular aperture field of view for
 both green and red-doublet emissions. 
But the calculated O($^1$S) and O($^1$D) production rates along the radial distances 
  are decreased by an order of magnitude above 10$^4$ km. Since OH is the dominant O-bearing 
species in the outer coma, the cross sections can affect the calculated the surface brightness
of [OI] 6300 \AA\ at larger projected distances ($>$10$^5$ km). To fit the observed [OI] 6300 \AA\
emission in the outer coma \cite{Glinski04} found it necessary to increase
theoretical determined OH to O($^1$D) photorate by a factor of around 3.

The chemistry model developed by \cite{Glinski04} suggested  that the collisions of O($^3$P) with 
OH leads to the formation of O$_2$. These calculations also showed that the O$_2$ densities can be as high 
as 1\% of H$_2$O. We evaluated the change in green and red-doublet emission intensities
 by incorporating O$_2$ in the model by taking its density profiles from \cite{Glinski04}. 
 No significant change ($<$5\%) is found
 in the green and red-doublet emission intensities by including O$_2$ 
in the model. This is because the other O-bearing species are  
several orders  of  magnitude higher in the inner coma.

\subsubsection{Effect of slit dimension on the derived O($^1$D) production rate}
As a case study, for a fixed H$_2$O production rate and CO and CO$_2$ relative abundances, 
we calculated [OI] 6300 \AA\ emission intensity over a
projected field of view for different slit dimensions. We then  derived the 
 O($^1$D) production rate based on the calculated average [OI] 6300 emission intensity over the 
projected field view. These calculations are presented 
in Table~\ref{tab-slt}. 
Since our model calculations are limited up to the projected distances of 10$^5$ km (which is discussed 
in Section~\ref{sec-lim}) we present the calculated intensities of [OI]  6300 and 5577 \AA\ emissions
 for the slit dimension up to  10$'$ $\times$ 10$'$.
Though O($^1$D) is substantially produced in the inner coma via photodissociation,
 the collisional quenching by cometary species results in a very few [OI] 6300 \AA\
 photons. The role of quenching in determining the [OI] 6300 \AA\ flux can be understood
from the calculated values presented in Table~\ref{tab-slt}. 
A large aperture observation is required, which covers the entire [OI] 6300 \AA\ emission region,
to derive the H$_2$O production rate. The 
calculations presented in  Table~\ref{tab-slt} suggest that by using large aperture 
slit the derived water production rate is closer to the actual production rate of H$_2$O.
 Hence, to derive the water production 
rate using [OI] 6300 \AA, the slit dimension which covers a projected distance more than the 
scale length of H$_2$O should be used.

\subsection{Limitations and future scope of the model}
\label{sec-lim}
The density of the species  produced in the inner coma (radial distances less than 10$^5$ km) 
is mainly controlled by photochemical reactions. Above these distances the transport of species 
starts becoming significant in determining the number density of the calculated species. 
Our model calculations
 are based on photochemical equilibrium condition and is for a collisional coma. Hence, 
model results presented at distances beyond 5 $\times$ 10$^5$ km are 
not as reliable as the values in the inner coma.
 Moreover, above these 
radial distances the chemical lifetimes of  neutral species are significantly altered  
by the solar wind interaction through charge exchange and impact ionization processes. 
Also, we could not incorporate altitude distribution of 
dust density in our model calculations which can affect the calculated optical depth.
 Since our model is time independent and one dimensional it is difficult to explain the
asymmetry in the observed [OI] 6300\AA\  emission intensity over the cometary coma. 
For determining the spectral width of green and red-doublet lines elaborated 
calculations are required along with laboratory measured photodissociation cross sections.

 \section{Summary and Conclusions}
 We have recently developed a coupled chemistry-emission model for the forbidden visible 
emissions 5577 and 6300 \AA\ of atomic oxygen in 
  comet \hyak\ \citep{Bhardwaj12}. In the present paper we applied our model to a high ($\sim$30 
times more 
than on \hya) gas production rate comet \hale\ in which these prompt emissions are observed 
in 1997 by \cite{Morgenthaler01} and \cite{Zhang01}. 
 The main results of our model calculations  on comet Hale-Bopp are  summarized as follows.
 \begin{enumerate}
 \item Below cometocentric distance  of 10$^3$ km, photodissociation of CO$_2$ is the 
 major production mechanism of O($^1$S). Between 10$^3$ and 10$^4$ km, the contributions from 
the photodissociation of CO$_2$ and H$_2$O are nearly equal.
  Above  2 $\times$ 10$^4$ km several other  processes are also 
significant to the O($^1$S) production. 

 \item Mainly  the solar 
photons in 955--1165 \AA\ wavelength band contribute to  the production of O($^1$S)
 in photodissociation  of CO$_2$. This is because the yield of O($^1$S) in CO$_2$ 
photodissociation reaches a maximum  in this wavelength region.

\item Since the cross section of photodissociation of CO$_2$ for the production of O($^1$S)
 is more than two orders of magnitude larger than that of H$_2$O, even a small amount 
(few percent relative abundance)  of CO$_2$ can make 
it an important source of the O($^1$S).
  
  \item Quenching by H$_2$O is the main loss mechanism for O($^1$S)  at radial
 distances below  300 km; above 
10$^3$ km  radiative decay via  5577 \AA\  emission is the dominant destruction mechanism. 
  
 \item Inside 10$^5$ km, the main production mechanism of O($^1$D) is photodissociation of 
H$_2$O; but, in the innermost part of the coma ($<$100 km) the photodissociation of CO$_2$ 
is also a significant source. 

 \item For photodissociation of H$_2$O, the peak O($^1$D) production  occurs 
 via H Ly-$\alpha$ (1216 \AA), 1165--1375 \AA\ and 1375--1575 \AA\ wavelength bands at  cometocentric 
distances of 1000, 200, and 50 km,
 respectively. Solar photons at all other wavelengths produce O($^1$D) with one or more orders 
of magnitude smaller efficiency. 
 
 \item Below 100 km, solar photons in the wavelength band 1375--1585 \AA\ mainly produce 
O($^1$D) by photodissociation of CO$_2$. The contribution from other wavelength bands is significant 
above cometocentric distances of 200 km. 

 \item The major destruction mechanism of O($^1$D) up to 3000 km cometocentric distance  
is quenching by H$_2$O; above 5000 km radiative decay takes over. 
  
  \item In comet \hbop\ the O($^1$D) density peaks occurs between 10$^3$ and 
10$^4$ km, while for O($^1$S) the peak is around 500--1000 km.

 \item The radiative efficiency of O($^1$S) and O($^1$D) atoms in comet \hbop\ are 
unity above 10$^3$ and 10$^4$ km, respectively. In comet \hya\ these distances are 10$^2$ and 
10$^3$ km, respectively.

\item The model calculated green to red-doublet emission intensity ratio is consistent with  
  the observation of \cite{Zhang01}.

\item Collisional quenching can change the G/R ratio by an order of magnitude, whereas 
 doubling the relative abundance of CO$_2$ increases its value by maximum of 30\%. 

\item To accurately measure the H$_2$O  production rate in cometary coma, a
 slit dimension which covers a projected distance more than the scale length of H$_2$O is 
preferred to cover the entire [OI] 6300 \AA\ emission region.
  
\item The model calculated [OI] 6300 \AA\ emission intensity profile as a function of projected distance 
 is in agreement with the observation of \cite{Morgenthaler01}. The model calculated  surface 
brightness averaged over a 4$'$ circular aperture field of view is  higher by a 
factor of 1.5 to 2 compared to the observation.

 \item The calculated mean excess velocity of O($^1$D) and O($^1$S) atoms in the region of 
 unity radiative efficiency  is $\sim$1.6 and $\sim$2 km s$^{-1}$, respectively, 
 which is consistent with the range of 
velocities observed by \cite{Cochran08} in several comets. 

 \item  Based on our model calculations for comets  \hya\ and \hbop,
 we conclude that [OI] 6300 \AA\ emission is mainly controlled by the 
photodissociation of H$_2$O, while the [OI] 5577 \AA\ emission line is 
contributed by both H$_2$O and CO$_2$. Since O($^1$S) 
production is associated with different molecules, whereas the O($^1$D) production
 is mainly from H$_2$O, the width of the green line will be higher than that of 
the red-doublet lines. 

 \end{enumerate}

With a high H$_2$O production rate, comet \hbop\  provided a large gaseous environment, 
which has not been seen in previous comets. 
Since the apparition was at small geocentric distances, the giant cometary coma has
provided a laboratory for investigating several collisional-driven
effects. These collision driven processes are very important in determining 
the distribution of cometary excited species in the coma, which manifests into the emissions 
of the cometary coma.

\section*{Acknowledgements}
 S. Raghuram was supported by the ISRO Senior Research
Fellowship during the period of this work.
 Solar Irradiance Platform historical irradiances are provided courtesy of W. Kent Tobiska and
Space Environment Technologies. These historical irradiances have been developed with
partial funding from the NASA UARS, TIMED, and SOHO missions. The authors thank  
the reviewers for their valuable comments and suggestions that has improved the paper significantly.

\end{document}